\documentclass[manuscript]{aastex}
\usepackage[utf8]{inputenc}
\usepackage{color}
\usepackage{lastpage}
\usepackage{ulem}

\newcommand{\eg}{{\it e.g.}}
\newcommand{\ie}{{\it i.e.}}

\newcommand{\PSone}{\protect \hbox {Pan-STARRS1}}

\begin{document}

\title{No Activity Among 13 Centaurs Discovered in the \PSone\ Detection Database}

\author{Eva Lilly\altaffilmark{1}}
\email{elilly@psi.edu}

\and

\author{Henry Hsieh\altaffilmark{1}, James Bauer\altaffilmark{2}, Jordan Steckloff\altaffilmark{1,3}, Peter Jev\v{c}\'ak\altaffilmark{4}, Robert Weryk\altaffilmark{5}, Richard J. Wainscoat\altaffilmark{5}, Charles Schambeau\altaffilmark{6}}

\altaffiltext{1}{Planetary Science Institute, 1700 E. Fort Lowell, Suite 106 Tucson, AZ 85719-2395, USA}
\altaffiltext{2}{University of Maryland, College Park, MD 20742-2421, USA}
\altaffiltext{3}{University of Texas at Austin, Austin, TX, 78712, USA}
\altaffiltext{4}{Comenius University, Mlynsk\'a dolina, 842 48 Bratislava, Slovakia}
\altaffiltext{5}{Institute for Astronomy, University of Hawaii, 2680 Woodlawn Drive, Honolulu, HI 96822, USA}
\altaffiltext{6}{University of Central Florida, 4000 Central Florida Blvd, Orlando, FL 32816, USA}

\begin{abstract}
Centaurs are small bodies orbiting in the giant planet region which were scattered inwards from their source populations beyond Neptune. Some members of the population display comet-like activity during their transition through the solar system, the source of which is not well understood. The range of heliocentric distances where the active Centaurs have been observed, and  their median lifetime in the region suggest this activity is neither driven by water-ice sublimation, nor entirely by super-volatiles.

Here we present an observational and thermo-dynamical study of 13 Centaurs discovered in the Pan-STARRS1 detection database aimed at identifying and characterizing active objects beyond the orbit of Jupiter. We find no evidence of activity associated with any of our targets at the time of their observations with the Gemini North telescope in 2017 and 2018, or in archival data from 2013 to 2019.
Upper limits on the possible volatile and dust production rates from our targets are 1-2 orders of magnitude lower than production rates in some known comets, and are in agreement with values measured for other inactive Centaurs.

Our numerical integrations show that the orbits of six of our targets evolved interior to r$\sim$15 AU over the past 100,000 years where several possible processes could trigger sublimation and outgassing, but their apparent inactivity indicates their dust production is either below our detection limit or that the objects are dormant. Only one Centaur in our sample -- 2014 PQ$_{70}$ experienced a sudden decrease in semi-major axis and perihelion distance attributed to the onset of activity for some previously known inactive Centaurs, and therefore is the most likely candidate for any future outburst.  This object should be a target of high interest for any further observational monitoring. 
\end{abstract}



\section{Introduction}
    Centaurs are small bodies most commonly defined as having perihelia ($q$) and semimajor axes ($a$) between the orbits of Jupiter and Neptune, i.e., 5.2 AU~$<q, a<$~30 AU \citep[\eg][]{Jewitt2009}, which are migrating towards the inner Solar system from their source regions in trans-Neptunian space. Once a trans-Neptunian object evolves to within Neptune's orbit and becomes a Centaur, it enters a chaotic, gravitational pin-ball machine where it is rapidly passed between the gravitational influences of the giant planets. Centaurs eventually leave the region via several routes - they collide with one of the planets, they return back to the trans-Neptunian region or even get ejected from the solar system, or about 20\% of Centaurs enters an end state as a Jupiter-family comet \citep[JFC;][]{Volk2008,Sarid2019}. Due to the chaotic gravitational environment, the typical dynamical lifetime of a Centaur is only $\sim$10 Myr \citep{Tiscareno2003,Disisto2007}. In this way, the Centaur population delivers useful clues of the pristine composition and physical properties of small trans-Neptunian objects (TNOs) beyond our current observational limits right to our doorstep. 

\begin{figure}[htbp]
    \centering
    \includegraphics[width=\textwidth]{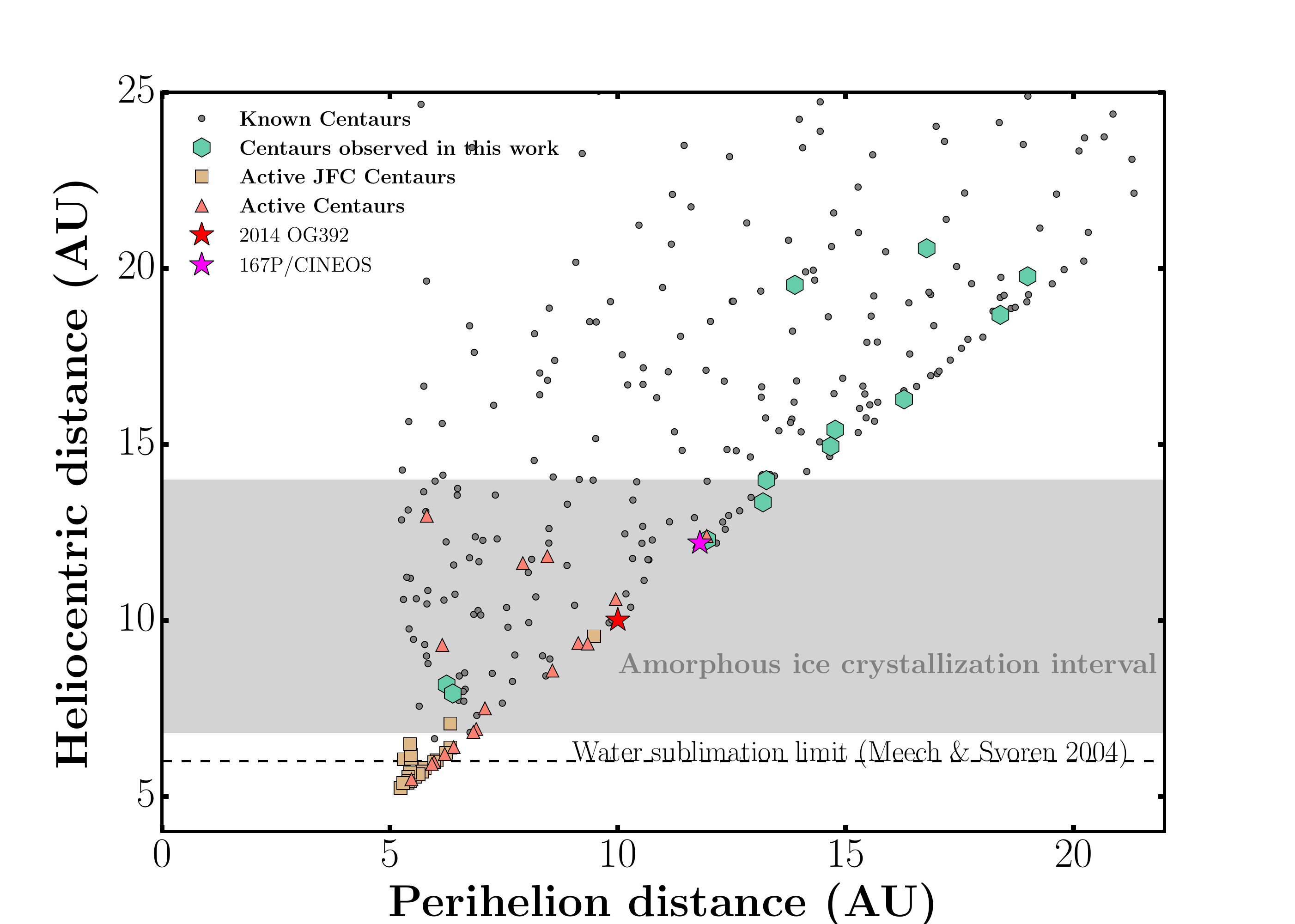}
    \caption{Perihelion distance vs. heliocentric distance of 13 Centaurs studied in this work compared to known active Centaurs and JFCs Centaurs, and the known Centaur population. All known active Centaurs reside within the interval where the crystallization of AWI is possible. Our targets are distributed both within and beyond the crystallization interval. A magenta star marks the position of 167P/CINEOS which shares a similar orbit and heliocentric position at the time of activity as one of Centaurs in our sample, 2014 OX$_{393}$.}
    \label{fig.Targets}
\end{figure}

One of the most puzzling characteristics of Centaurs is comet-like activity exhibited by $\sim$5\% of the population \citep{Chandler2020}. Unlike in most short-period comets, this activity is not water-driven because the majority of active Centaurs resides at large heliocentric distances where the surface temperatures are too low to allow for water ice sublimation \citep{Meech2004,Womack2017}. 
Active Centaurs also show a variety of behavior, from long-term low-level outgassing such as in (2060) 95P/Chiron \citep{Luu1990,Foster1999}, to sudden violent outbursts in previously known dormant objects, as observed in 174P/Echeclus \citep{Choi2006,Bauer2008,Rousselot2008,Kareta2019,Seccull2019}, and unlike in regular comets, the timing of the activity onset is not obviously centered on perihelion.  Dust production rates often range by orders of magnitudes from one object to another, and the source of this activity remains unknown. Most known Centaurs have spent the majority of their lifetime in region too warm to maintain surface supplies of super-volatile ices such as CO and CO$_2$ \citep{Jewitt2009}, for example if free CO ice was present on all Centaurs, we should observe Centaurs with comae well beyond the orbit of Neptune, just like in some long period comets \citep[\eg][]{Szabo2008,Meech2017b}, which is not the case. Multiple works have shown there are no known active Centaurs beyond r$\sim$14~AU, and that the lack of activity cannot simply be attributed to observational selection effects \citep{Jewitt2009,Cabral2019,Li2020}. The fraction of active objects among Centaurs with perihelia smaller than 14~AU is almost 20\%, while only ~10\% of Centaurs are active within the overall population. Moreover, almost 27\% of all active Centaurs has perihelia smaller than 6.5 AU, as can be seen on Figure~\ref{fig.Targets}. 

Therefore considering the wide range of heliocentric distances at which Centaurs exhibit comae, the Centaur activity likely has more sources than just sublimation of surface volatiles, and these sources may be triggered by multiple mechanisms, some of which could be different than those acting in regular comets.

It has been suggested that activity on Centaurs and some comets at large heliocentric distances is driven by the crystallization of amorphous water ice (AWI), which is porous and can contain molecules of super-volatile ices trapped in the AWI during the condensation \citep{Notesco1996,Jewitt2009,Shi2015}. When the parent body enters a warmer environment, a crystallization front starts propagating into the interior of the body, releasing trapped gas molecules which then start to loft dust grains at temperatures much lower than necessary for sublimation of crystalline water ice. AWI has been detected by in situ observation of the Galilean moons of Jupiter, \citep{Hansen2004}, however, the presence of AWI has never been observationally confirmed on small bodies in the Solar system\citep{Lisse2013}.  

The phase transition of AWI to crystalline water ice has been proposed to start at a heliocentric distance of r$\sim$7~AU \citet{Jewitt2009,Womack2017}, while recent thermal modeling studies show it can be triggered up to r$\sim$~14 AU, with some extreme cases crystallizing as far as at  r$\sim$~16 AU \citep{Guilbert2012}. Figure~\ref{fig.Targets} shows that the known active Centaurs indeed have perihelia within this heliocentric distance range, which could support the idea of AWI crystallization as their main activity driver.  However, Figure~\ref{fig.Targets} also shows that there are other apparently inactive bodies with very similar orbital parameters. Thus, the mere presence of AWI, which should be abundant in TNOs and Centaurs \citep{Jenniskens1994}, does not appear to be enough on its own for an object to exhibit activity, indicating that activity may require a certain trigger. \citet{Fernandez2018} suggested that a rapid change in the thermal environment of the body could jump-start AWI crystallization in a dormant object. Such an orbital change could be caused by a close encounter with a planet and subsequent decrease of perihelion or semi-major axis, which is believed to have led to outbursts in several Centaurs and the previously dormant quasi-Hilda P/2010 H2 (Vales) \citep[\eg][]{Mazzotta2006,Fernandez2018,Jewitt2020}. 

Another plausible explanation for increased activity in Centaurs that have been only recently scattered inward would be the sublimation of subsurface pockets of CO$_2$ and other super-volatile ices.  CO$_2$ in particular is significantly more volatile than water ice, and is one of the most common species found on cometary bodies \citep[\eg][]{Bockelee2004,AHearn2011}. Thermal models show that it can sublime vigorously interior to $\sim$15~AU \citep{Steckloff2015,Meech2017}, which coincides closely with the heliocentric distances of observed Centaur activity. Such subsurface deposits could be forcefully exposed by some external force and start to rapidly sublimate. We can rule out impacts as the predominant surface activity trigger on Centaurs due to the very low collision rate in the region \citep{Durda2000}, but other forces might play a role. For example, \citet{Steckloff2016a} has shown that topographic features on the surface of comets, when steeper than $\sim$30 degrees, are prone to mass wasting events that can excavate buried volatiles such as CO and CO$_2$ to the surface.  

An important aspect to consider is also the time an object has already spent in the Centaur region, or the past changes in its orbit pushing its orbit within the AWI crystallization limit or past the point where the sublimation of other supervolatiles is possible. In general it is expected the active Centaurs have young orbits - \ie they have only recently entered the region, since the timescale on which the crystallization front is able to propagate before it depletes the AWI supplies is fairly small (in order of 10$^5$ years, \citet{Guilbert2012}), and also the depletion of the CO-layers is rapid. \citep{Li2020}. 

 CO has been detected in the comae of several Centaurs \citep[\eg][]{Womack1997,Wierzchos2017, Womack2017}, while CO$_2$ that is notoriously challenging to detect with ground-based assets, has only been observed in comets so far \citep{Ootsubo2012,Womack2017}. It is unclear if the observed production rates were caused by gases released during the AWI phase transition or from sublimation of subsurface supervolatile ice pockets. Therefore  to better understand the types of activities displayed in the Centaur region we need to gain a more informed and complex view of the orbital and thermal histories of Centaurs and the current status of their volatile content.

Observations for this work have been collected in the framework of a long-term program aimed at searching for active Centaurs and analyzing their orbital and thermal environment. Here we present the analysis of our pilot study on a sample of 13 Centaurs that were discovered in the detection database of the Panoramic Survey Telescope and Rapid Response System survey \citep[\PSone ;][]{Weryk2016}.


\newpage
\section{Methods}
\label{s.Methods}
In this work we have observed and analyzed 13 Centaurs discovered in the Pan-STARRS1 detection database \citep{Weryk2016}. All our targets have well-defined orbits with orbital arcs spanning almost a decade. Our investigation consisted of three steps: 1) analysis of images taken with  Gemini-N to search for comet-like activity, 2) a search of archival images from the Dark Energy Camera (DECam) to look for traces of activity in the past, 3) thermal modeling to examine surface temperatures as a function of orbital parameters and object size to put upper limits on sublimative gas production from volatile species potentially present on the surface, and 4) numerical integrations to inspect the orbital histories of objects in our sample.


\subsection{Observations and Data Reduction}
\label{ss.ObsDataRed}

We observed our targets with the GMOS-N (Gemini Multi-Object Spectrograph) instrument on the Gemini North telescope on Maunakea, Hawaii between 8/1/2017 and 1/31/2018 (\# GN-2017B-Q-69). GMOS-N has a 300 x 300 arcsec field of view and a pixel scale of 0.0728~arcsec~px$^{-1}$. For our observations, we used the imaging mode of GMOS N with 2x2 binning and a Sloan Digital Sky Survey (SDSS) r' filter \citep{Fukugita1996}, which is ideal for detecting dust \citep{Meech2004}.  We employed non-sidereal tracking at the apparent rate and direction of motion for each target, and acquired sequences of three 100~s exposures, which were necessary to correctly identify each target based on their motion.
All our observations were done in queue mode, meaning each object was observed under slightly different conditions throughout the semester. Our observations are summarized in Table \ref{tab.ObsConditions}.

Standard bias subtraction, flatfield correction, and cosmic ray removal were performed for all images using Python 3 code utilizing the {\tt ccdproc} package in Astropy\footnote{\url{http://www.astropy.org}} \citep{astropy2018_astropy} and the {\tt L.A.Cosmic} python module\footnote{Written for python by Maltes Tewes (\url{https://github.com/RyleighFitz/LACosmics})} \citep{vandokkum2001_lacosmic,vandokkum2012_lacosmic}. Images from each night were aligned on the object and co-added together to create composite images using IRAF software \citep{tody1986_iraf,tody1993_iraf} in order to maximize signal-to-noise (S/N). Absolute photometric calibration was carried out using a photometry pipeline developed by \citet{Mommert2017}, standard IRAF routines, and the all-sky ATLAS-REFCAT2 catalog \citep{Tonry2018}.

\subsubsection{Target Selection}
\label{sss.TargetSelection}

Our target list consisted of 13 Centaurs selected from a sample of 29 new Centaur discoveries from the Pan-STARRS1 detection catalog \citep{Weryk2016} that were observable in the second half of 2017 from the Northern hemisphere and were within Gemini-N observation limits. We included objects with perihelia spanning the entire Centaur region and crossing the putative AWI crystallization heliocentric distance interval \citep{Jewitt2009,Guilbert2012}. Most of the known active Centaurs were discovered close to perihelion and were active at discovery.  We therefore primarily selected targets close to perihelion to ensure the best conditions for observing possible activity.  As of June 2021 none of our targets have been subject of characterization by other works.

\begin{figure}[htbp]
    \centering
    \includegraphics[width=0.75\textwidth]{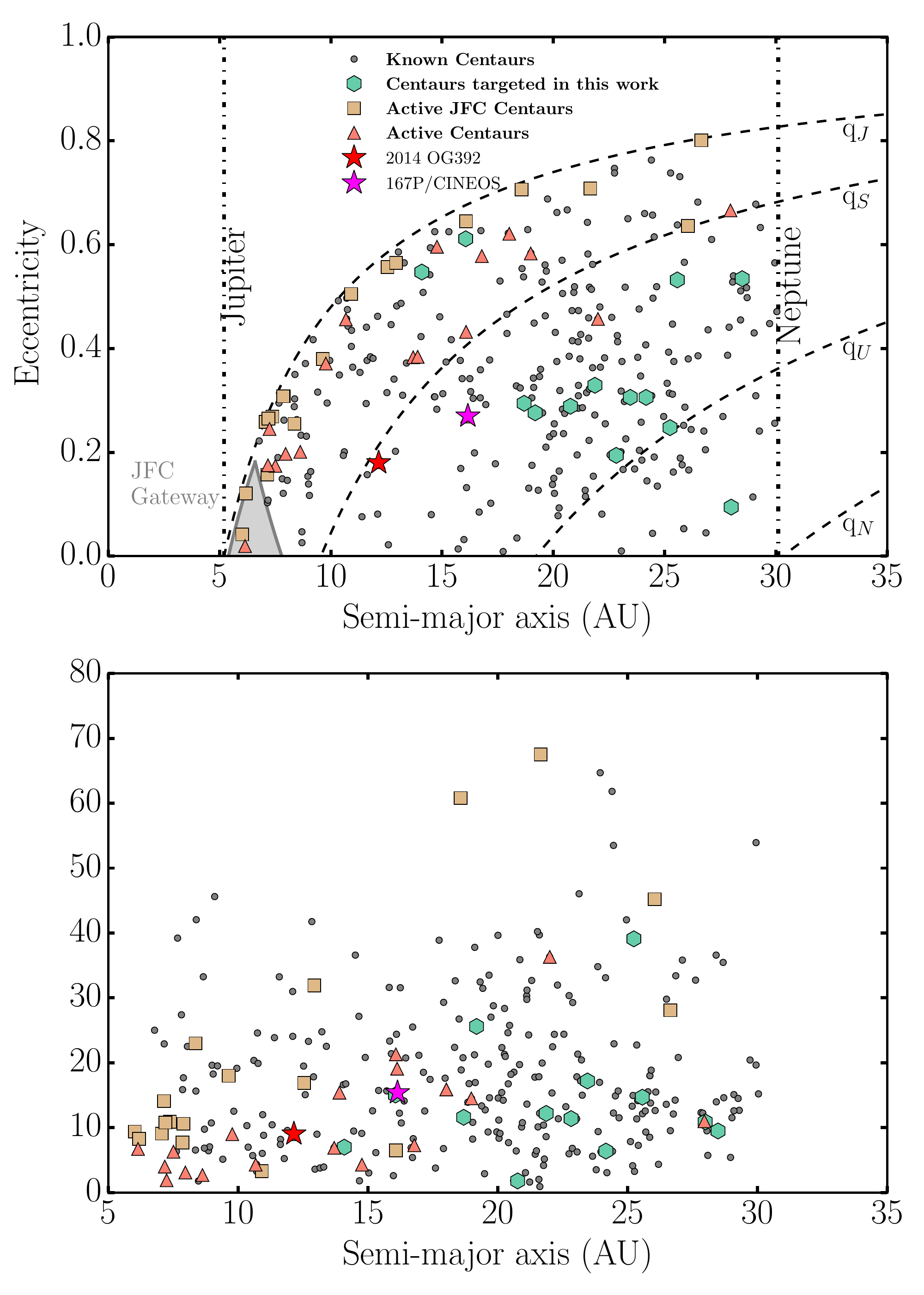}
    \caption{Semi-major axis versus eccentricity (top panel) and semi-major axis versus inclination (bottom panel) of our targets compared to known active Centaurs and JFC Centaurs, and the known Centaur population. Dashed curves represent the loci of orbits with perihelia equal to those of perihelia of giant planets, dashed-dotted lines show semi-major axes of Jupiter and Neptune which border the Centaur region. Solid grey triangle represent the JFC Gateway region according to \citet{Sarid2019}.}
    \label{fig.Targets-a-vs-e}
\end{figure}

Figures~\ref{fig.Targets} and ~\ref{fig.Targets-a-vs-e} show that our targets cover a wide range of orbital elements typical for Centaurs, and stay clear of the region often referred to as the ``JFC Gateway'', a temporary low-eccentricity region exterior to Jupiter through which the majority of Centaurs bound to become JFCs pass through \citep{Sarid2019,Steckloff2020}. In contrary, most objects in our sample have perihelia near the upper limit of the AWI crystallization zone because the data-mining algorithm applied to the \PSone\ detection database was tuned to find objects beyond Neptune and the discovered Centaurs were mostly a byproduct of that search \citep{Weryk2016}.  It can also be seen that the activity in Centaurs does not seem to be related to any preferential eccentricity or inclination ranges, but instead their perihelion distances appears to be the main constraint.

By definition, JFCs orbiting in the same region as the Centaurs (\ie, 5.2 AU~$<q, a<$~30 AU) differ only by the fact they are active, and
in terms of Tisserand's parameter with respect to Jupiter, $T_J$, which is given by
\begin{equation}
    T_J = a_J/a + 2\cos i \sqrt{a/a_J (1 - e^2 )}, 
\end{equation}
where $a_J$ is the semimajor axis of Jupiter, and (a,e,i) is the semimajor axis, eccentricity, and inclination of the Centaur respectively.
JFCs have 2$> T_J >$3 and perihelia below $\sim$6~AU, while Centaurs typically have larger T$_J$ and larger perihelia, signaling that their orbits are decoupled from Jupiter \citep{Fernandez2018}. To add to the confusion, the Gateway objects are by definition Centaurs, but their orbits are not decoupled from Jupiter, while there also are inactive Centaurs with $T_J<$3 in the population. Objects with perihelia between 5.2 AU and 6 AU are thus on the border of these two classes and they often make trips to within Jupiter's orbit and back to the Centaur region - the typical example being the famous Gateway Centaur  2019 LD$_2$ \citep{Steckloff2020},. Therefore we use the term ``JFC Centaurs'' throughout this paper for addressing these borderline active objects. There are a total of 18 active Centaurs and 21 JFCs Centaurs known in the region as of 2021 February 10\footnote{https://physics.ucf.edu/~yfernandez/cometlist.html\#ce}.

\begin{figure}[htbp]
    \centering
    \includegraphics[width=\textwidth]{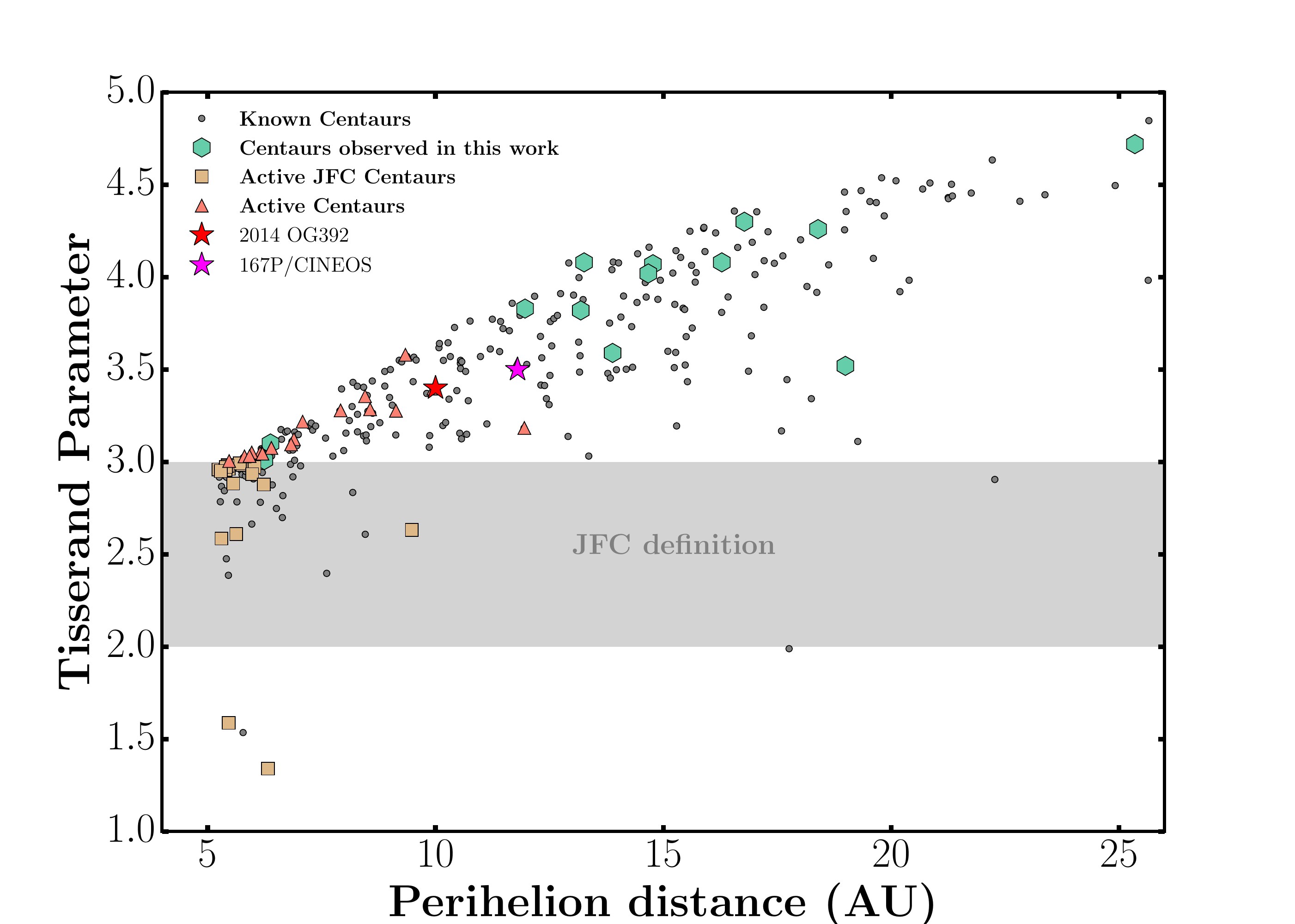}
    \caption{Perihelion distance vs. Tisserand Parameter with respect to Jupiter of 13 Centaurs studied in this work compared to known active Centaurs and JFC Centaurs, and the known Centaur population as of February 2021.}
    \label{fig.Targets-T-vs-q}
\end{figure}

We have also calculated Tisserand's parameter with respect to Saturn and Uranus - $T_S$ and $T_U$ respectively, and  Table~\ref{tab.Orbparameters} shows that about half of our targets are likely more influenced by the outer planets than Jupiter. The current values of $T_J$ of our targets and other bodies in the region are plotted against their perihelion distance, and are shown in Figure~\ref{fig.Targets-T-vs-q}.  
Again, we do not note any apparent correlation between activity and $T_J$.

\subsection{Archival Search}
\label{sss.ArchivalSearch}

We used the Canadian Astronomy Data Centre's Solar System Object Image Search \citep{gwyn_2012_pasp} to search the Dark Energy Survey (DES; \citealt{abbott_2018}) archive for serendipitous observations of our 13 Centaurs from the Dark Energy Camera (DECam; \citealt{decam_2015_aj}) installed on the Cerro Tololo Inter-American Observatory's Blanco 4-meter telescope. The DECam instrument consists of 62 individual 2k$\times$4k CCDs for imaging in a hexagonal arrangement on the focal plane. This results in a 2.2 degree diameter field of view and 0$''$.263 per pixel plate scale. The combination of a relatively large field of view and the deep imaging afforded by the Blanco telescope results in a treasure trove of serendipitously imaged Solar System small bodies using DECam. Such a search yielded the  recent discovery of an active Centaur, 2014 OG$_{392}$ \citep{Chandler2020}.

\begin{figure}[htbp]
    \centering
    \includegraphics[width=0.8\textwidth]{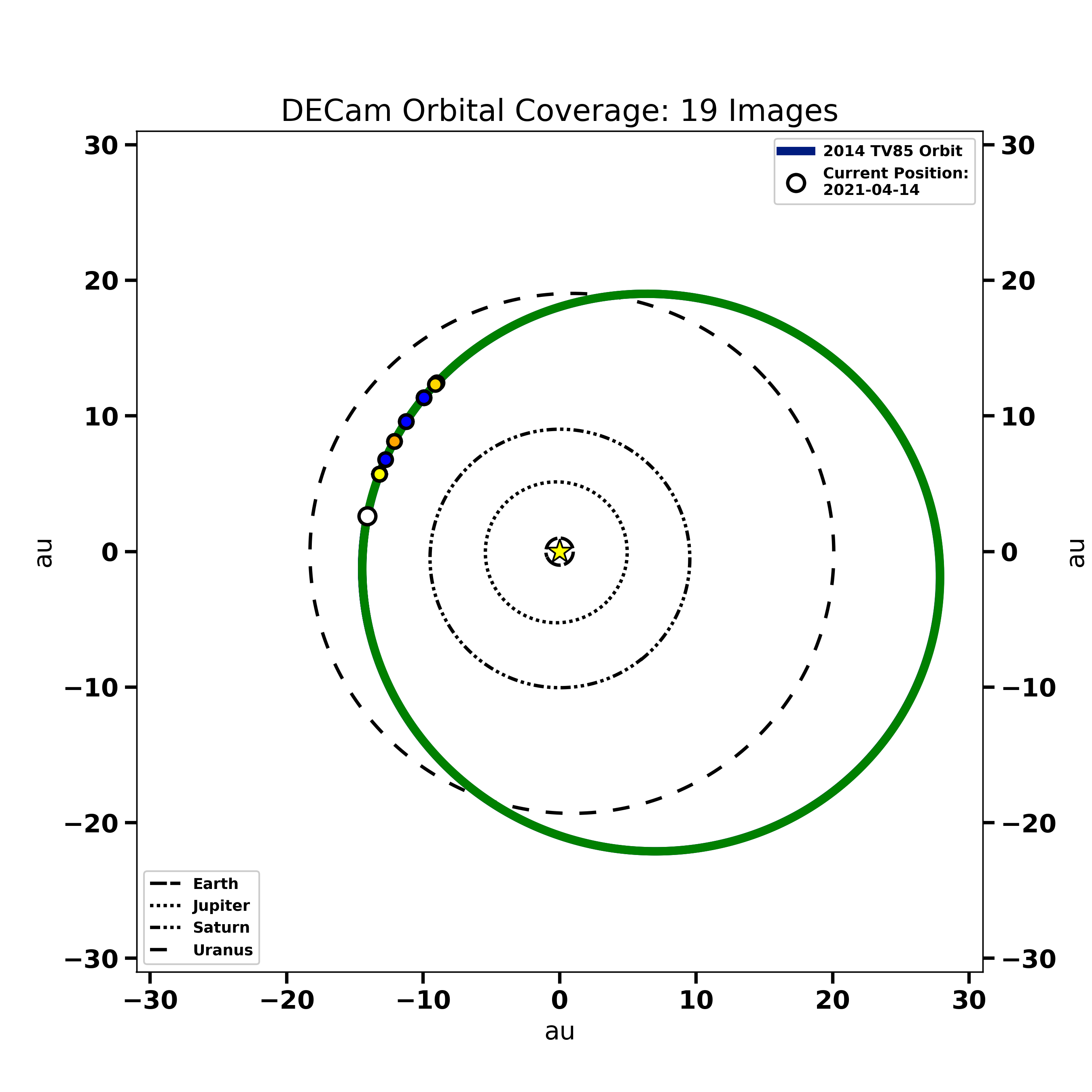}
        \caption{The orbital image coverage of 2014 TV$_{85}$ from the DECam archival images. The colored circles on the Centaur’s orbit indicate the orbital locations for which there are DECam images. There were a total of 19 archival DECam images of the object, out of which 8 were suited for calibration. The calibrated data span 4 years (2014 -- 2018).}
    \label{fig.DECam_data}
\end{figure}

Since the archival data we used only cover observation epochs between $\sim$2013 and 2019, they do not provide significant orbital coverage of a typical Centaur orbit, as their orbital periods are several decades- to several hundred years long. However, these archival data are still useful as snapshot observations to search for possible activity.
For reference, Figure~\ref{fig.DECam_data} shows the orbital arc of one of our Centaur targets (2014 TV$_{85}$) that is covered by data from DECam.

Our DES search revealed numerous epochs of data for all but three of our targets (2010 TU$_{191}$, 2014 PQ$_{70}$ and 2014 WX$_{508}$).  DECam's archived multi-extension FITS images have undergone basic reductions including bias subtraction, flat field correction, and fitting of a World Coordinate System as described in \cite{morganson_2018}. We also applied our Python-based DECam image processing pipeline to individual images to: (1) identify the FITS extension containing coverage of each target based on its JPL Horizons ephemeris coordinates, (2) save the identified individual CCD data and header metadata as a new FITS image file, (3) perform cosmic ray removal utilizing the LACosmic technique \citep{van_dokkum_2001} as implemented in ccdpro \citep{matt_craig_2017}, and (4) perform photometric calibration based on Pan-STARRS field stars \citep{Tonry2018}.


\section{Results and Discussion}
\label{s.Results_Discussion}

\subsection{Photometric analysis and Nuclei sizes}
\label{ss.Activity}

\begin{figure}[htbp]
    \centering
    \includegraphics[width=\textwidth]{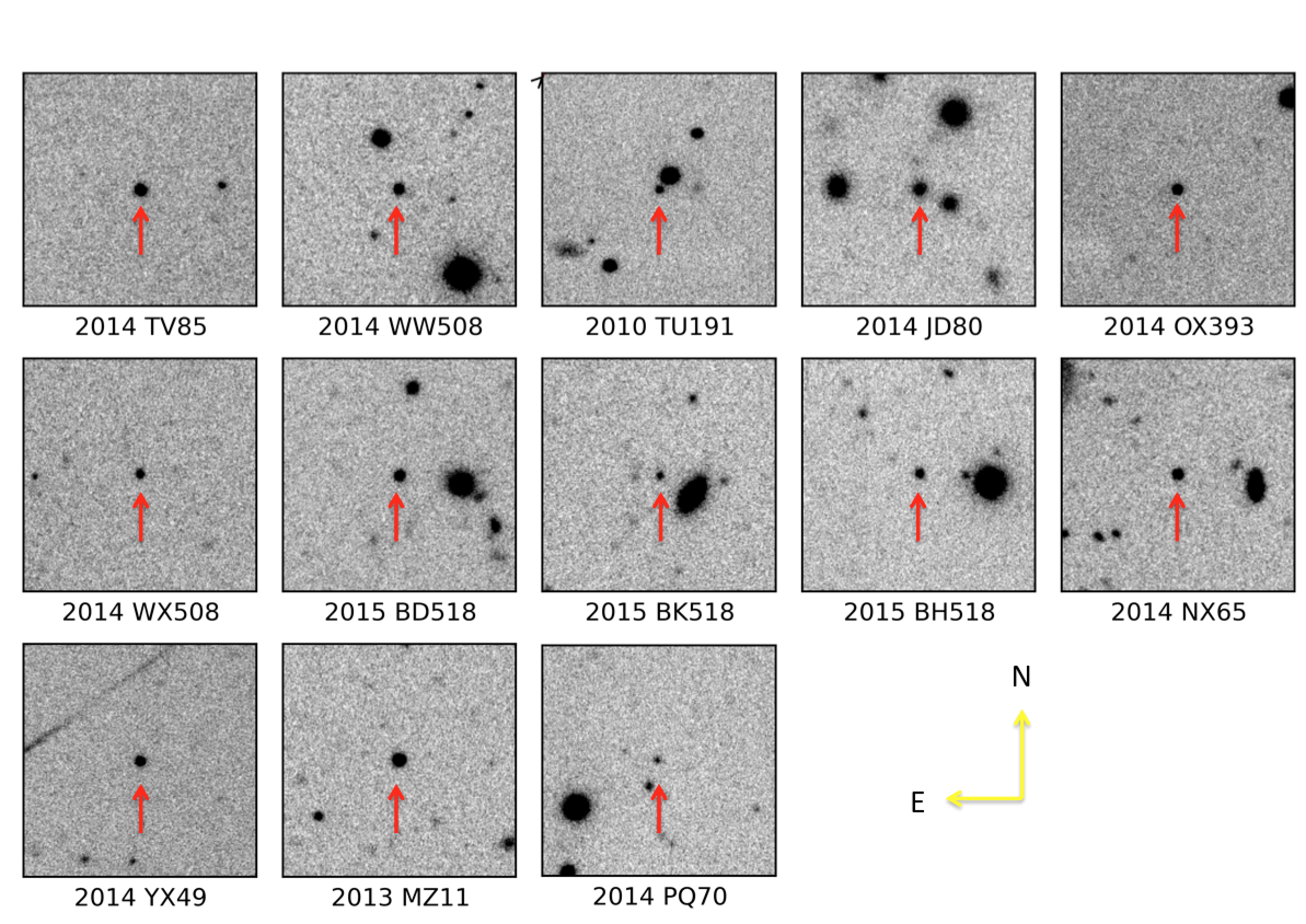}
    \caption{Postage stamp cutouts from stacked images of 13 Centaurs observed in this work. The spatial dimensions of each stamp are 30 x 30 arcsec and each stamp is centered on the object with North up and East to the left. None of the objects showed obvious visual signs of activity, conclusion supported by the SBP analysis (see Sec.~\ref{ss.Activity}).}
    \label{fig.Stamps}
\end{figure}

In order to search for activity in our observational data, we first identified the correct object in each star field by ``blinking'' the individual images, and visually inspecting each one for extended surface brightness features moving along with the source. We then stacked the images using standard IRAF routines to increase the S/N. All our targets appeared stellar based on simple visual inspection, as can be seen on Figure~\ref{fig.Stamps}.

\begin{figure}
\centering
\includegraphics[width=\textwidth]{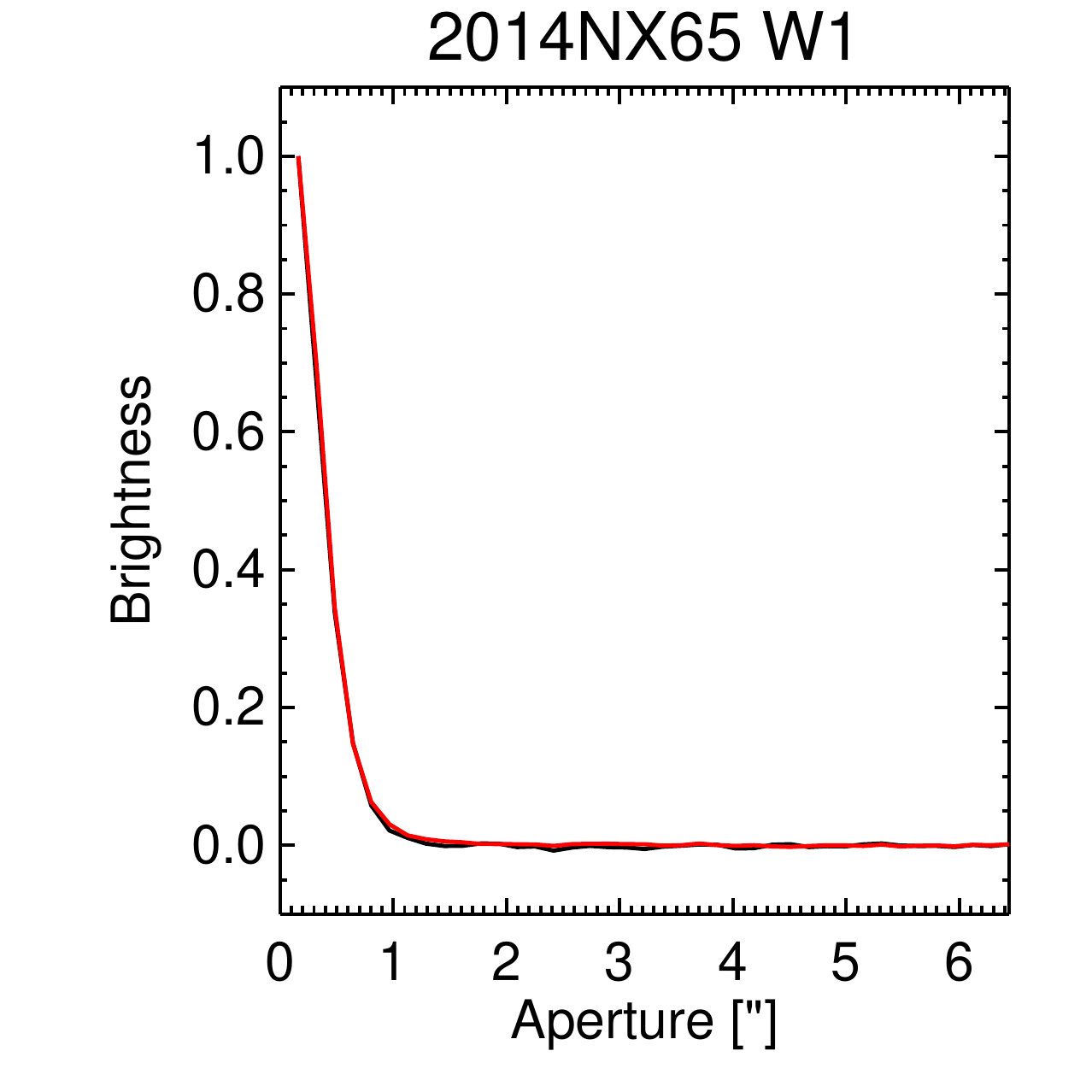}
\caption{Radial surface brightness profile (SBP) of  2014 NX$_{65}$ (black line) compared to the average profile of background field stars (red line). Both profiles match almost exactly, and therefore rule out faint activity of our target down to the surface brightness of 24.0 mag~arcsec$^{-2}$.}
 \label{fig.SBP-2014NX65}
\end{figure}

To reveal activity much fainter than could be discerned by a visual inspection, we calculated surface brightness profiles (SBPs) of each target using the method of \citet{Meech1997} and compared it to the average SBPs of nearby stars.  SBP analysis is a powerful method for detecting faint coma, and by averaging
the mean brightness values within a fixed distance
from the Centaur nucleus we can increase our sensitivity
to coma many-fold. Fortunately our targets were at heliocentric distance large enough that their sky-plane motion was less than 0.1``/min, and the trailing of the stars used in the SBP comparison was therefore negligible. In the most extreme case of the fastest moving Centaur in our sample - 2014 NX$_{65}$ the tracking elongation was about 21\% of the seeing FWHM values and hence undetectable. SBP analysis further revealed no evidence of low level activity in any of the Centaurs in our sample (e.g., 2014 NX$_{65}$; Figure~\ref{fig.SBP-2014NX65}). 


We calibrated our photometric data using sources from the GAIA \citep{Gaia2016} and ATLAS REFCAT2 catalogues \citep{Tonry2018} as photometric references, and converted the apparent r'-band magnitudes, $m_r$, measured for our targets to absolute r'-band magnitudes, $H_r$ (i.e. apparent magnitudes normalized to $r = \Delta = 1$~AU and a solar phase angle of $\alpha=0^{\circ}$) using: 
\begin{equation}
    H_r = H_r(\alpha) + 2.5\log[(1-G_r)\cdot \Phi_1(\alpha) + G_r\cdot \Phi_2(\alpha)], 
    \end{equation}
where the reduced magnitude, $H_r(\alpha)$, i.e., the apparent magnitude normalized to $r=\Delta=1$~AU, is given by
    \begin{equation}
    H_r(\alpha) = m_r(r,\Delta,\alpha) - 5\log(r\Delta)
    \end{equation}
where $m_r(r,\Delta,\alpha)$ is the apparent magnitude at a given heliocentric distance $r$, geocentric distance $\Delta$, and phase angle $\alpha$ at the time of observation, with the average IAU slope-parameter $G = 0.13\pm0.12$ measured in R-filter for Centaurs by \citet{Bauer2003}. The $\Phi_i$ functions are given by 
\begin{equation}
\Phi_i(\alpha) = \exp\left\{- A_i\left(\tan\frac{\alpha}{2} \right) ^{B_i}\right\}
\end{equation}
where $A_1$ = 3.33, $A_2$ = 1.87, $B_1$ = 0.63 and $B_2$ = 1.22. 

We then used our calculated absolute magnitudes and the average Centaur geometric albedo of $p=0.08$ measured by \citep{Bauer2013} from the infrared WISE data to calculate effective nuclei diameters for our objects. The resulting values are summarized in Table~\ref{tab.Mag_radii_production}. However, Centaurs are a unique small body population given their striking color bimodality and a wide span of albedoes \citep[\eg][]{Peixinho2004,Peixinho2012}, where the  measured mean albedoes for the two distinct Centaur groups are  6$\pm$2\% for the blue group, and 12$\pm$5\%  for the redder color group \citet{Bauer2013}. Therefore, if our targets belong to the blue group, they would be 15.4\% larger on average, and if they are in the red group, they would be 18.4\% smaller than if they had the mean albedo.

The photometric measurements reveal that our sample consists of an ensemble of average-sized Centaurs with effective radii of $2.7<R<60$~km if we use the mean albedo, where 2014 PQ$_{70}$ is one of the smallest Centaurs ever observed with a mean nucleus radius of 2.7 km, similar to a typical JFC \citep{Bauer2003,Fernandez2013}.
    
Table~\ref{tab.DECam_obs} summarizes the archival images from DECam that contain Centaurs in our sample and were suitable for photometric calibration, along with observing geometry parameters, filters used for the observations in question, and measured apparent photometric magnitudes. These images were taken for other research projects, and therefore they were taken using a variety of filters and exposure times. Out of 13, only 3 objects, 2015 PQ$_{70}$, 2010 TU$_{191}$ and 2014 WX$_{508}$, were not present in the archive. For the rest of our targets, we have applied the SBP analysis following the procedure applied to the Gemini data, which revealed no traces of a faint coma on the archival images similar to what we have observed with Gemini.

Three objects from our target list --- 2014 JD$_{80}$, 2014 NX$_{65}$ and 2013 MZ$_{11}$ --- were also observed with the Hubble Space Telescope (HST) 
by \citet{Li2020}, who searched for activity among high-perihelion objects crossing the Centaur region. None of the HST targets has exhibited signs of activity.


\subsection{Thermal environment and its effect on activity levels}
\label{ss.Thermal_evolution}

In our analysis of the thermal environment of the inspected Centaurs we were focusing on two possible activity sources: The solar-driven sublimation of surface volatiles and the crystallization of AWI capable of releasing trapped gasses. Multiple volatile species have been observationally detected in active Centaurs \citep[\eg][]{Senay1994,Womack1997,Wierzchos2017,Womack2017}, while there is no direct evidence of the presence of AWI \citep{Lisse2013} on Centaur surfaces. It has to be noted that due to their distance and their corresponding faintness, the detection of any volatile species on Centaurs sized similarly to our sample is extremely challenging \citep{Kareta2021}.

To asses the plausibility of activity driven by sublimation of surface volatile deposits, we have computed the theoretical maximal sublimation rates for the three most common volatiles found in comets, CO, CO$_2$, and H$_2$O for each of our targets at their observed heliocentric distances. We used the thermal model described by \citet{Steckloff2015}, which assumes a spherical graybody covered entirely in the pure volatile species of interest, with an albedo of approximately 4 percent, and emissivity of 0.9. The model calculates the dynamic sublimation pressure and the sublimative mass loss rate across the surface of a planetary body by numerically solving for the equilibrium temperature that balances incident solar energy with radiative and sublimative heat loss. The model integrates over the entire surface of the spherical nucleus and accounts for lower sunlight intensity toward the terminator due to increasing solar incidence angles, assuming all sunlight is used to sublime volatiles.  This model thus estimates maximal production rates capable from solar-driven activity.

Assuming comet-like dust-to-gas mass ratios $f_{dg}$, the calculated maximum sublimation rates can then be used to compute an idealized maximal dust production rate of the body under the assumption the whole surface is active.

Our simplified modeled sublimation rates are summarized in Table~\ref{tab.Mag_radii_production} and assume only pure CO, CO$_2$, or H$_2$O surface ice as the major producer of gas molecules. Our results show the CO-sublimation rate is many orders of magnitude larger than CO$_2$ and H$_2$O-sublimation rates for our targets at their respective heliocentric distances at the time of observation. CO thus has the strongest volatility in the
region we are inspecting and therefore we will consider it as the most likely sublimating ice species, and we will use it to constrain the active surface area and the dust production rates on our targets throughout the rest of this paper.  

It is unrealistic to expect that CO or other supervolatiles would cover the whole surface at the analyzed heliocentric distances --  we would observe strong surface CO-driven activity much farther away from the Sun because CO starts to sublimate at 25K. It also has been shown \citep{Wierzchos2017} that known Centaurs with detected CO-emissions (with the one excepion of 29P/Schwassmann-Wachmann 1 (29P) \citep{Womack2017}) are actually significantly CO-depleted compared to comet Hale-Bopp, which could be considered an unprocessed nucleus and had similar size and produced CO molecules at similar heliocentric distance. However, objects that may appear depleted in volatile species due to the lack of detected activity  could still retain sizeable sub-surface volatile deposits.

Idealized models of cometary nuclei depicting them as porous spheres composed of a mixture of dust and ices \citep{Capria2000,Capria2002} have shown that both in the cases of pure CO ice and trapped CO gas, the source material for CO-driven activity is likely stored under the surface where the temperatures drop quickly, and in theory Centaurs should continuously release CO along their orbits due to its low sublimation temperature and the shallow depth of the deposits \citep{Mazzotta2006}, which has not been observed. It is possible that during the residence in the Centaur region objects gradually undergo surface devolatilization and develop a dusty crust inhibiting the heat transfer and gas permeability of upper layer effectively reducing the sublimation rate and gas flow to the surface \citep{Wierzchos2017}. In the case of CO$_2$, most of the volume is likely stored just under the surface, but distant inactive Centaurs might have retained some surface patches of CO$_2$ since this supervolatile starts sublimating at r$\sim$15~AU \citep{Steckloff2015,Meech2017}. However, the CO$_2$ has not been detected on Centaur surfaces yet, because it is even more challenging than detecting CO with our current facilities, although it has been measured in situ on comets \citep{AHearn2011}.

Therefore unless a Centaur is dynamically young in the region and still retains surface CO/CO$_2$ deposits, or has not developed dust crust yet, we can expect the CO/CO$_2$ sublimation from the subsurface to be relatively weak and possibly go undetected until an external or internal force disturbs the volatile-rich layers.

If subsurface pockets of amorphous water ice are present, crystallizing, and releasing trapped volatiles to the surface, they could contribute to the gas flow from the sub-surface. Laboratory experiments have shown trapped gas could indeed be relatively abundant: AWI has the ability to retain gases up to a gas-to-ice molecule ratio $f_{GI}=3.3$ \citep{Laufer1987}. 
However it should be noted that the crystallization of AWI is not capable of heating pockets of volatiles from below the surface because the phase transition cannot be both exothermic (releasing heat) AND releasing trapped volatiles. \citet{Kouchi2001} showed that AWI absorbs heat when it crystallizes and releases trapped volatiles for concentrations of impurities (\eg\ CO or CO$_2$ molecules) above ~2\%, which are typical in the interstellar dust and comets. The AWI phase transition itself therefore cannot produce enough heat to cause rapid sublimation of subsurface volatiles resulting in outburst, but it can act as a force opening the way to the surface and to provide additional gas drag. For example the well known active Centaur 29P has been reported to have CO and dust outbursts that are not correlated and could be associated with multiple release mechanisms acting in parallel \citep{Wierzchos2020}.

All above implies the apparently inactive/dormant Centaurs orbiting within the AWI crystallization zone have probably already depleted their surface volatiles, since the average CO sublimation loss in the region is several meters per orbit \citep{Li2020}, and the CO$_2$-driven activity would be visible, but they could still contain subsurface pockets of volatiles and AWI buried under the dust crust. A sudden rapid orbital change pushing the perihelion closer to the Sun could jump-start the crystallization front, which will move rapidly releasing trapped gas until enough pressure builds up to cause a landslide or opens a sinkhole exposing subsurface layers. This process could lead to large-scale outgassing and/or outburst, possibly with a significant lag, as was observed in several active Centaurs \citep{Mazzotta2006} and also in quasi-Hilda P/2010 H2 (Vales) \citep{Jewitt2020}. Such rapid release of gasses could also eject a large boulder or a chunk of the parent body which can continue to disintegrate, as has been likely observed during a dramatic ourburst of 174P/Echeclus in 2005, where the the nucleus and coma brightness peaks were spatially separated \citep{Rousselot2008,Bauer2008,Kareta2019}. As the crystallization front moves through the body, it could keep building the pressure pockets and create openings throughout the surface even far away from perihelion, and in subsequent orbital passages until the volatile deposits in each pocket are depleted.

However, crystallization cannot be sustained for a time longer than 10$^4$–10$^5$ years \citep{Guilbert2012} further suggesting that active Centaurs have young orbits, and have only recently been scattered inside to the warmer regions of the Solar system. 

Using  174P/Echeclus as an example, we can calculate the volume of AWI necessary to release the observed dust volume. Given the dust production rate observed during the 2005 outburst - 400 kg~s$^{-1}$ \citep{Bauer2008}, $f_{GI}$ for AWI, the density of water ice ($\rho_{\rm H_2O}=1000$~kg~m$^{-3}$), and the range of dust-to-gas ratios measured for comets and estimated for active Centaur 29P - $f_{dg}=$1-8 \citep[Y. Fern{\'a}ndez, personal communication;][]{Fernandez2020,Wierzchos2020}, it would require between $\sim$~15 to 120~kg of AWI to crystallize every second and release trapped CO to match the observed dust production if AWI is the single source. A $\sim$~1 month long outburst would require between $10^7 - 10^8$~kg of amorphous ice, or a sphere of $\sim$~20-40~m in diameter, which is an AWI pocket similar in size to what \citet{Jewitt2020} calculated for a Hilda asteroid undergoing an outburst of similar magnitude.

It should be noted that our calculated values should only be used as an order-of-magnitude estimates given there are no direct simultaneous measurements of gas and dust mass loss from an active Centaur. The f$_{dg}$ ratio can change depending on the Centaur and whether the mass loss is a product of a quiescent activity, or an outburst \citep{Fernandez2020}. A fresh sinkhole opening up and causing an outburst could be much dustier than a steady-state outgassing event diffusing through surface regolith. For example \citet{Wierzchos2020} has shown that in case of 29P several CO outbursts did not have a corresponding dust outbursts. There is currently a real lack of knowledge of the gas-to-dust ratio of Centaurs due to observational limitations, which will be hopefully addressed and alleviated with future works using JWST \citep{Kelley2016}, but such rigorous analysis is beyond the scope of this paper.

\subsection{Mass loss rates and Fractional Active Surface area}
\label{ss.Mass_loss}

We used the maximum CO sublimation production rates from our thermo-dynamical model to derive the maximum surface sublimation-driven dust production rate,  $dM_{CO}/dt$, assuming CO to be the most likely volatile to drive the activity at heliocentric distances of our targets, and $f_{dg}=$1 estimated for Centaur 29P during its quiescent activity phase \citep{Fernandez2020,Wierzchos2020}. This value most likely represents the volume of dust expected to be lofted by steadily escaping gas from a weakly-active Centaur such as targets of this study. The in situ $f_{dg}$ measurements on 67P revealed the ratio being much higher closer to the nucleus - $f_{dg}=$4$\pm$2, compared to ratios $f_{dg}=$0.1-2.0 typically measured remotely for comets\citep[\eg][]{Weiler2003,Choukroun2020}, but the comet had an active water ice sublimation, and was near its perihelion at the time the measurement were taken, thus it could be expected the volume of dust being lofted was higher than in our targets.

\begin{figure}[htbp]
    \centering
    \includegraphics[width=\textwidth]{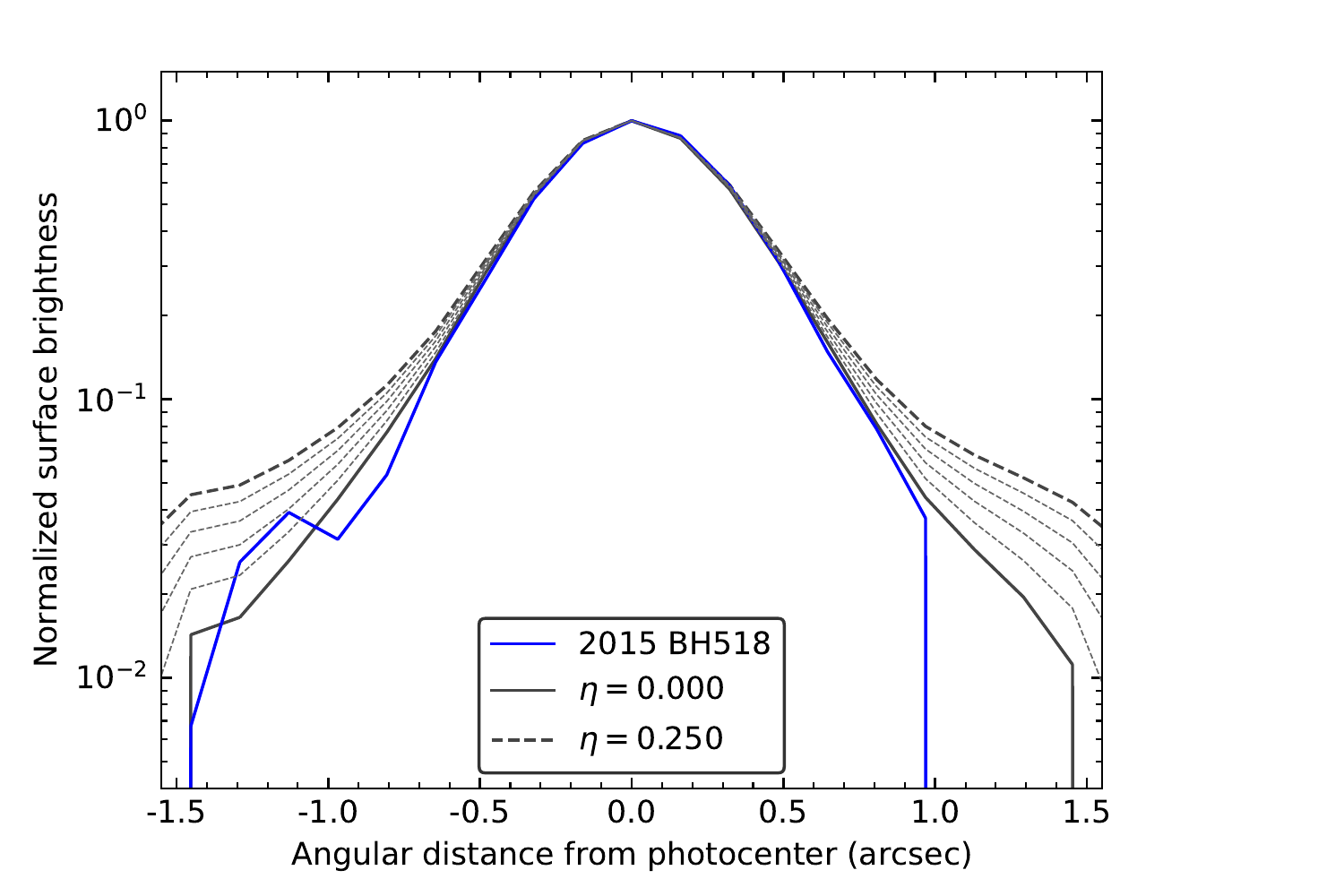}
    \caption{Comparison of the PSF for object 2015 BH$_{518}$ with a series of synthetic PSFs with varying coma levels $\eta$. The coma parameter for this object was 0.05.}
    \label{fig.coma_model}
\end{figure}

 
Then using the synthetic coma limit technique following \citet{Luu1992}, we calculated the observed effective upper limit on the dust production rate. To find quantitative limits on weak activity, we compared every Centaur's PSF with seeing-convolved models of weak coma. With the same image scale and point-spread function as our data, models and data can be directly compared in order to ascertain activity levels. We first created model point sources, then generated a series of synthetic PSFs with varying coma levels and convolved these with the seeing (Figure~\ref{fig.coma_model}). Coma levels were parameterized by $\eta=C_c /C_n$, where $C_c$ and $C_n$ are the scattering cross-sections of the coma and nucleus, respectively, and the reference photometry aperture radius used was $\Phi=4.04$~arcsec.  For our analyses, we used coma levels of $\eta=0.00$ to $\eta=0.25$ in $\eta=0.05$ increments.

Increasing coma levels generated this way generally has little effect on the profile cores but does broaden the profile wings. We then noted at which coma level were the synthetic profiles most similar to the profile of the target object. Our comparison of the coma models to measured profiles of our sample Centaurs produced no indication of activity within our detection limits. However, we used the resulting limiting coma parameter $\eta_{lim}$ to convert to the observed effective upper limit on the dust production rate $dM_D/dt$ via:

\begin{equation}
    dM_D/dt = \frac{(1.1 \times10^{-3}) \cdot \pi~\rho~\bar{a}~ \eta_{lim}R^2}{\Phi~r^{0.5}\Delta},
\end{equation}
where $\rho=2500$ kg m$^{-3}$ is the assumed grain density consistent with carbonaceous chondrites, which
are associated with primitive C-type objects, and are most likely the closest match to the Centaurs \citep{Britt2002}, $\bar{a} = 0.5 \times 10^{-6}$~m is the assumed weighted mean grain radius \citep{Bauer2008}, $R$ is the  object's effective radius, $\Phi$ is the angular photometry radius in arcseconds, $r$ is the heliocentric distance in astronomical units, and $\Delta$ is the geocentric distance in astronomical units \citep{Luu1992}. 

By comparing the maximal production rates of CO and CO$_2$ derived from the surface sublimative mass-loss and the upper limit on dust production rates derived from our observations, we can calculate the effective fraction of active surface area of the particular Centaur at the time of the observation, and the actual active area as a function of object's diameter. Since we are using a photometric method to derive the upper limits on dust production, the possible changes in dust-to-gas ratio affect the size of the effective fraction of active surface area - the active area scales with $1/f_{dg}$, because a larger active area is needed to lift the same amount of dust grains at fixed gas flow, if the $f_{dg}$ is smaller.
The resulting production rates are in order of several kg with the effective active surface areas for each target ranging from several- to several-thousand m$^2$ in case of CO-driven activity. Our values are 1-2 orders of magnitude smaller than values calculated for other weakly active Centaurs \citep[\eg][]{Bauer2003,Mazzotta2014,Mazzotta2017}. Since the gas production rates from a CO$_2$-driven activity are several orders of magnitude lower than those calculated for CO, and are strongly dependent on the heliocentric distance, the active surface areas constrained by our coma-limit technique would be unrealistically large for our most distant targets. However, patches only several hundred  m$^2$ of sublimating surface CO$_2$ would sustain the observed dust production limits on two of our closest targets - 2014 PQ$_{70}$ and 2015 BK$_{518}$.

If this mass-loss volume is to be attributed to an escaped gas from AWI crystallization, it would require a relatively modest subsurface pocket akin to a sphere with several-m radius, which is in agreement with AWI pocket sizes estimated for outbursting quasi-Hilda P/2010 H2 (Vales) \citep{Jewitt2020}. Resulting values are summarized in Table~\ref{tab.Mag_radii_production}.

Our dust production upper limits can be well explained by a combined gas drag from volatiles released from crystallized AWI, and from sublimating ice patches under the surface exposed to the sun by the pressure of escaping molecules.

Cometary nucleus models \citep[\eg][]{Prialnik1995,Capria2000} predict that CO molecules should percolate through the outer layers of the body and flow uniformly from anywhere on the surface on both the night and day sides. This  indicates there should also be an uniform associated dust release due to the gas drag, unlike the dust mass loss caused by water sublimation which tends to be released only from the day side. Also, it has been proposed that sublimation of CO and CO$_2$ and the associated dust production rate changes very slowly with time after the activity onset because it originates in the subsurface layers \citep{Capria2000,Capria2002}. This would mean if there are large enough subsurface CO-deposits on Centaurs, most of the dynamically younger population members should exhibit a low-scale outgassing and dust mass-loss along their orbits similar to what has been observed in (2060) 95P/Chiron, which would be in most cases fainter than our detection limits \citep[\eg][]{Kareta2021}. 

\newpage
\subsection{Orbital evolution}
\label{ss.Orbital-evolution}

To track past orbital evolution and identify any recent potential activity-triggering events such as significant orbital changes or close encounters with giant planets,
we ran several sets of 
backward numerical integrations for each of our targets. We used the Bulrisch-St{\"o}er integrator in the Mercury6 N-body integration package \citep{Chambers1997}, which is capable of handling close planetary encounters with minimal loss of precision by using an adaptive timestep once an object crosses massive body's Hill sphere. 
All integrations used a 30-day timestep, which is an sufficiently precise to map the evolution of orbits with decades-long periods typical for Centaurs. Particles were removed from the integrations when they collided with the Sun or a planet, or reached semi-major axes of $a>100$~AU and were considered to be ejected from the solar system.
Dynamical clones were generated for each object using covariance matrices from the JPL Small Body Database  \footnote{https://ssd.jpl.nasa.gov/sbdb.cgi}. All of our targets have well defined orbits with up to 10 opposition arcs and low residuals resulting in small dispersion of clones from the nominal orbit.

For our first set of integrations, we produced 50 clones for each object, and ran them as massless particles under the gravitational influence of eight planets and the Sun for 100,000 years into the past, which is the the upper time limit on amorphous water ice survival on Centaurs \citep{Guilbert2012}. The orbits in the Centaur region are subjected to strong gravitational influence from giant planets and their orbital dynamics becomes chaotic very quickly. Therefore it is nearly impossible to know the system's dynamical evolution past the point in the history where the clones start to diverge \citep{Morbidelli2020}.

After examination of these long-term integrations we ran two more sets of integrations: a) using a time step of 10 days for past 5,000 years in order to investigate clone evolution before the individual orbits started to diverge, and b) using a 1-day timestep for 200 years into the past to search for extremely recent orbit changes similar to other Centaurs observed in outburst  \citep[\eg][]{Fernandez2018}.

\begin{figure}[htbp]
    \centering
    \includegraphics[width=\textwidth]{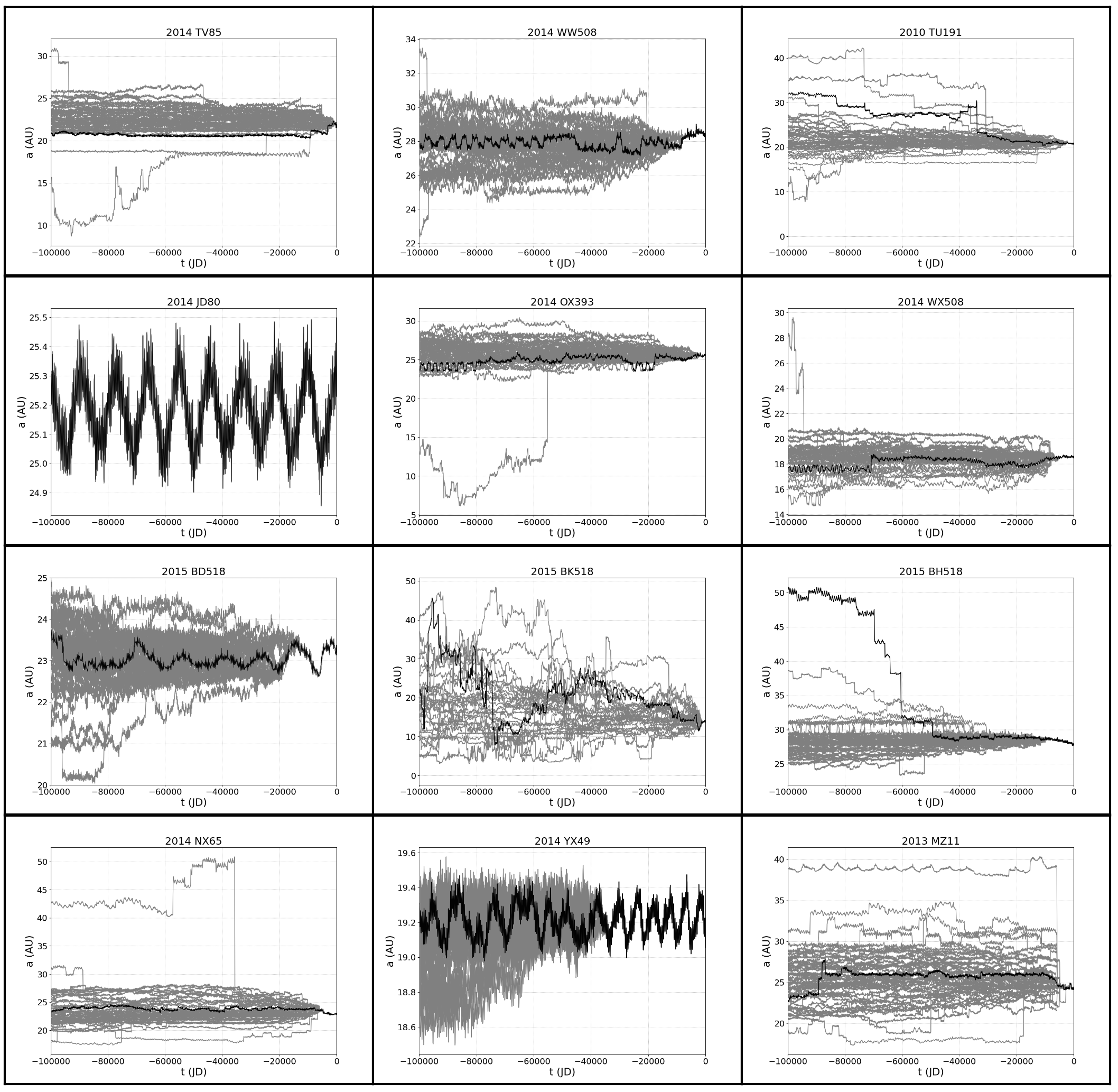}
    \caption{The evolution of semi-major axis of 12 Centaurs over the past 100,000 years.}
    \label{fig.Sma-evolution-100k}
\end{figure}

\begin{figure}[htbp]
    \centering
    \includegraphics[width=\textwidth]{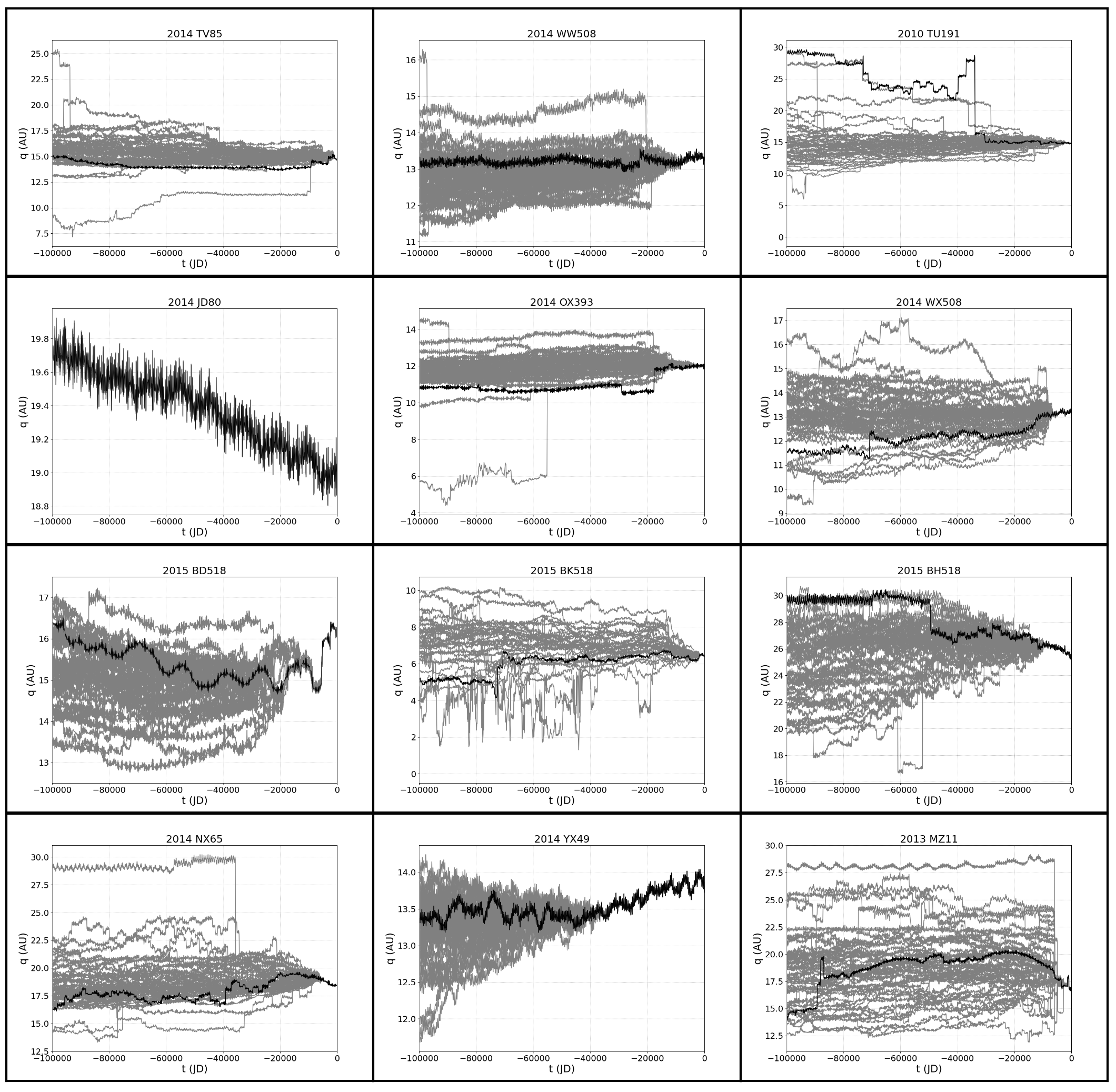}
    \caption{The evolution of perihelion distance of 12 Centaurs over the past 100,000 years.}
    \label{fig.Peri-evolution-100k}
\end{figure}

\begin{figure}[htbp]
    \centering
    \includegraphics[width=\textwidth]{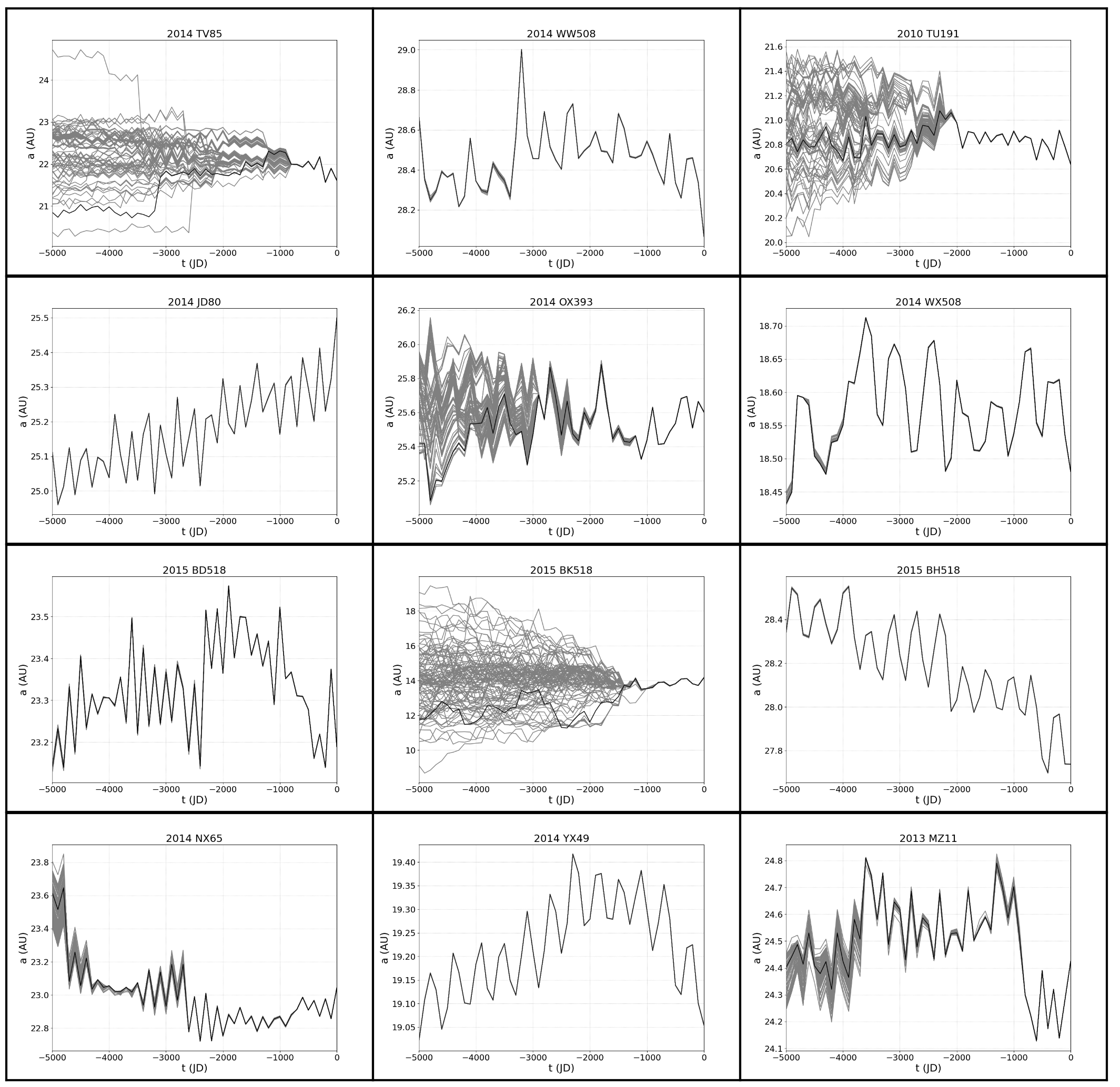}
    \caption{The evolution of semi-major axis of 12 Centaurs over the past 5000 years.}
    \label{fig.Sma-evolution-5kyr}Hearn
\end{figure}

\begin{figure}[htbp]
    \centering
    \includegraphics[width=\textwidth]{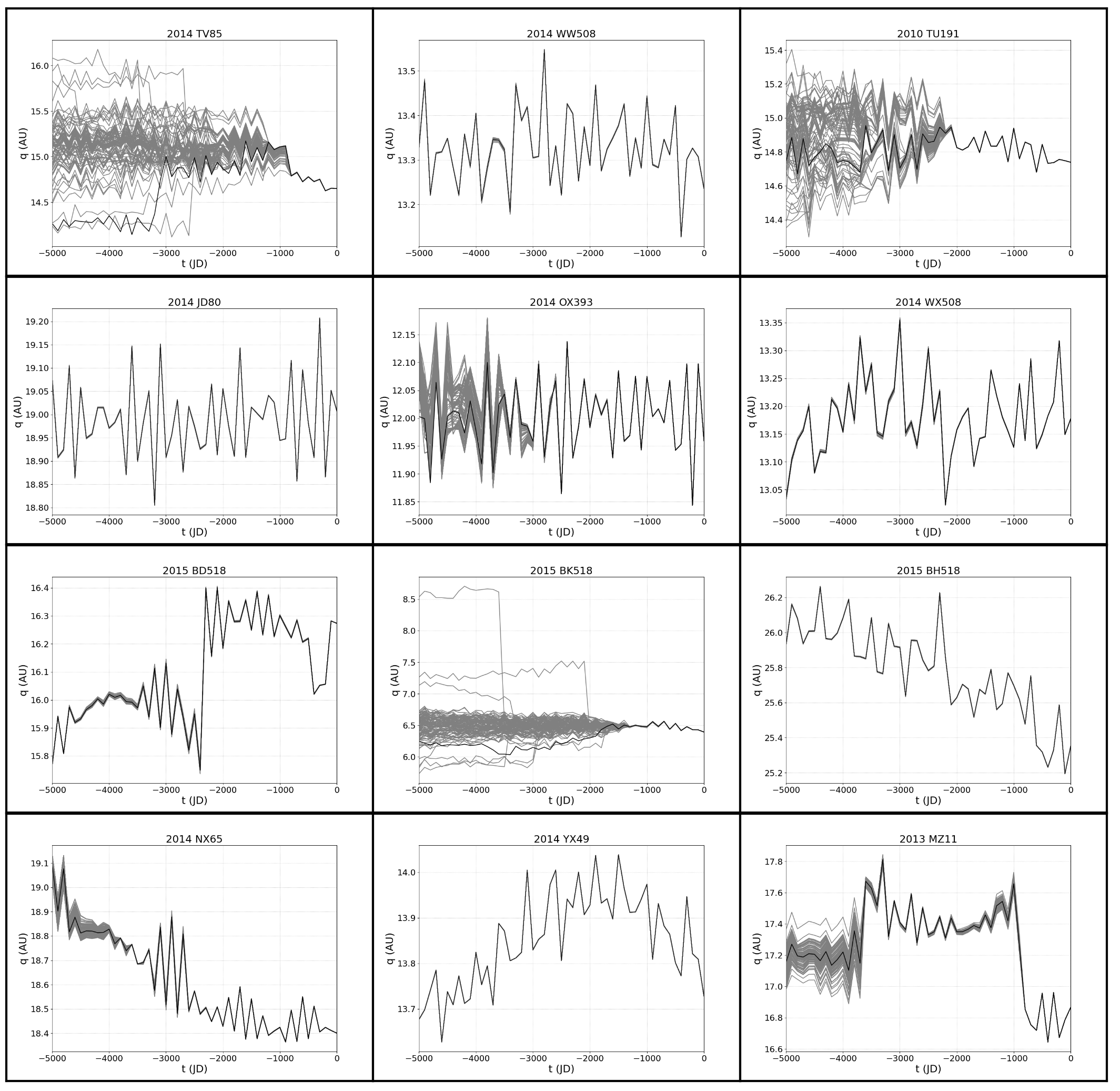}
    \caption{The evolution of perihelion distance of 12 Centaurs over the past 5000 years.}
    \label{fig.Peri-evolution-5kyr}
\end{figure}

Our analysis of the orbital histories revealed that most of the Centaurs in our sample have relatively stable orbits with only minor orbital changes as can be seen in Figures~\ref{fig.Sma-evolution-100k} and \ref{fig.Peri-evolution-100k}. Our 5 kyr integrations show that the nominal orbit of each target and those of their clones start to diverge on $\sim$10$^3$~yr timescales where the chaotic environment makes any determination of the evolution before this point statistical only, as shown on Figures~\ref{fig.Sma-evolution-5kyr} and \ref{fig.Peri-evolution-5kyr}. Such a point of divergence is typically associated with a close encounter with a planet.

\begin{figure}[htbp]
    \centering
    \includegraphics[width=\textwidth]{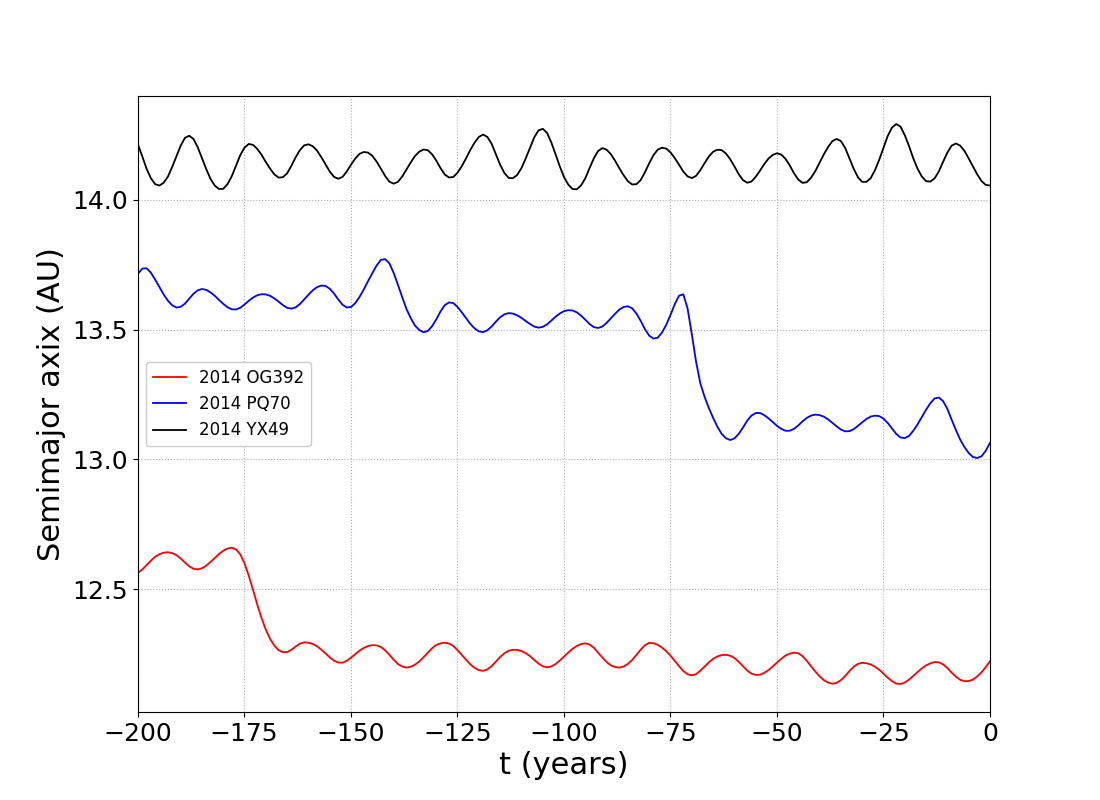}
    \caption{The evolution of semi-major axis of the recently discovered active Centaur 2014 OG$_{392}$, and two apparently inactive Centaurs observed in this work. Centaur 2014 PQ$_{70}$ has experienced decrease in semi-major axis of similar amplitude as 2014 OG$_{392}$ The semi-major axes of 2014 PQ$_{70}$ and 2014 YX$_{49}$ have been scaled down by 5 and 3 AU respectively for better comparison on the plot.}
    \label{fig.PQ70-sma}
\end{figure}

\begin{figure}
\centering
\includegraphics[width=0.75\textwidth]{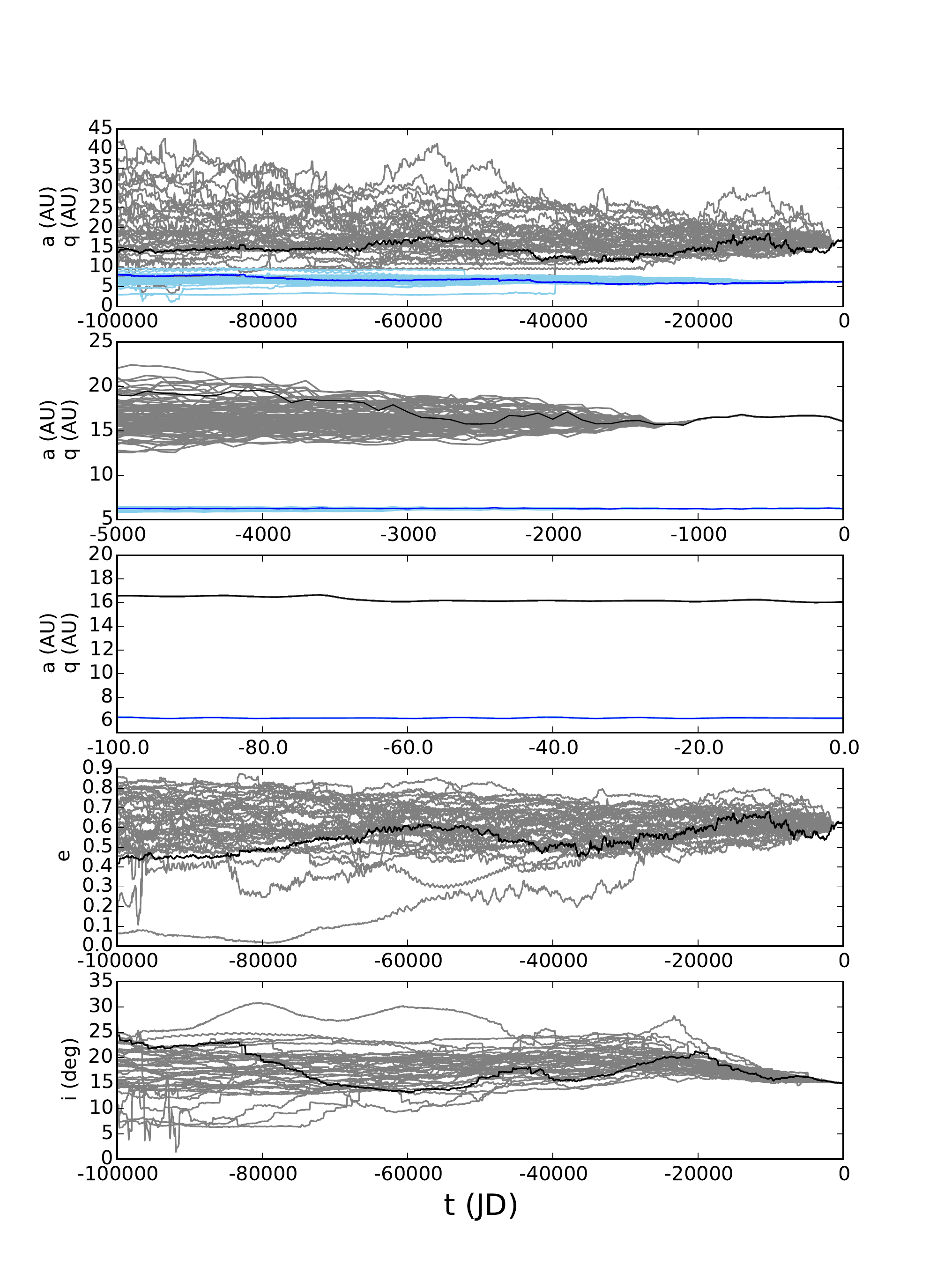}
\caption{Orbital evolution of low-perihelion Centaur 2014 PQ$_{70}$. Three upper-most panels show the orbital evolution of semi-major axis (black line) and perihelion distance (blue line) for 100 kyr, 5 kyr and 200 years into the past (top to bottom). Two bottom panels show the evolution of eccentricity and the orbital inclination in the last 100 kyr.}
 \label{fig.PQ70-ev}
\end{figure}

We can divide the observed orbital behavior of our targets and their expected activity levels into three groups based on perihelion distance:
\begin{enumerate}
    \item $q\gtrsim15$~AU: Objects with perihelia outside of the AWI crystallization zone (and CO$_2$ sublimation limit) tend to have stable orbits on the considered timescale, with changes in perihelion and semi-major axis of $\Delta a, \Delta q \lesssim 1-3$ AU. Objects 2015 TV$_{85}$, 2014 JD$_{80}$, 2015 BH$_{518}$, 2015 BD$_{518}$, 2014 NX$_{65}$, 2013 MZ$_{11}$ and most clones of 2010 TU$_{191}$ fall within this group. We expect very limited activity driven by surface or subsurface volatiles due to fairly long residence inside Neptune's orbit and therefore the depletion of surface sources, and we do not expect significant crystallization of AWI at the distances they reside in.
    
    \item 7 AU~$\lesssim$q$\lesssim$~15AU: Objects with perihelion distances in the AWI crystallization zone. Centaurs 2014 YX$_{49}$, 2010 OX$_{393}$, 2014 WX$_{508}$, and 2014 WW$_{508}$ belong to this group, and it is entirely possible the AWI crystallization/CO$_2$ has been underway, but the released gasses remain under the surface or remain below our detection limit. However, objects in this group experienced only minor orbital changes in the past 10$^5$~yrs, and we would expect only minimal gas-and dust loss driven by the crystallization since the front should have stopped propagating inside the bodies or has propagated too deep into the nucleus over the residence time in the zone \citep{Guilbert2012}. As we noted before, temperatures in this region are too high for any supervolatiles to have survived on the surface until present day.
    
    \item $q\lesssim7$~AU: There are two Centaurs in our sample with perihelia and semi-major axes below 7 AU: 2015 BK$_{518}$ and 2014 PQ$_{70}$. Both have current $T_J>3$ and $T_S<3$, and their orbital evolution can be traced with a reasonable precision $\sim2000$ yr into the past, where the orbits of clones start to turn chaotic under the strong gravitational influence from Saturn. Our integrations show wide diffusion of possible semi-major axes while the perihelia changed very little from the current value, similarly to what other works have found on the evolution of low-perihelion Centaurs \citep{Fernandez2018,Kareta2019,Steckloff2020}. It is possible that Centaurs in this group experienced periods of activity in the past, given their proximity to Sun, with surface volatiles completely depleted and perhaps some remaining pockets of subsurface ices still isolated under the surface. If they entered the AWI crystallization zone within last $\sim$10$^5$ years, the AWI crystallization front might be propagating under the surface, and with favorable conditions can become active again. For example 2014 PQ$_{70}$ in particular would be good candidate for an outburst monitoring campaign because it has experienced a decrease in semi-major axis $\sim$~70 years ago (Figures~\ref{fig.PQ70-sma} and~\ref{fig.PQ70-ev}) similar to what has been observed in other active Centaurs \citep{Mazzotta2006}, and we could witness yet another bout of activity triggered by this event in coming years. 
\end{enumerate}

There are three other notable Centaurs in our sample.
2010 TU$_{191}$ may be a fairly new addition to the Centaur region. Figure~\ref{fig.Peri-evolution-100k} shows that three of the clones including the nominal orbit experienced a close encounter with Neptune within 0.03 AU $\sim38,000$ years ago, reducing the perihelion distance by 15 AU within a decade and illustrating some drastic orbital changes that are common in the Centaur region. This object could represent a `fresh' Centaur with high likelihood of undisturbed surface volatile deposits, and it is entirely possible it is weakly active below our detection limits, and might become increasingly more active in the future.

Meanwhile, 2014 JD$_{80}$ is in a 3:2 mean-motion resonance with Uranus, and additional 0.5 Myr integration suggests it will stay resonant for next $\sim$1 Myr (K.\ Volk - personal communication).

Lastly, an apparently inactive object (2014 OX$_{393}$) that we observed in this work has a very similar ratio of perihelion distance and heliocentric distance at the time of observation, and orbital elements as two known active Centaurs --- 167P/CINEOS (248835) and 2006 SX$_{368}$ \citep{Jewitt2009,Jewitt2015} --- at the times of their outbursts (Figure~\ref{fig.Targets}), which could indicate 2014 OX$_{393}$'s inactivity could be due to having a different orbital history compared to the other two. While 2014 OX$_{393}$ has a stable orbit just inside the upper limit of the crystallization zone, it either lacked an activity trigger in the form of a sudden perihelion drop and rapid warming of the surface, or the past activity has depleted  surface volatiles.

With one exception, our fully-inactive Centaur sample has had relatively stable orbits, adding support to the idea that sudden orbital changes might trigger or re-start activity in the Centaurs.

\section{Conclusions}
\label{s.Conclusions}

We have observed 13 Centaurs with the Gemini North telescope with the goal of searching for comet-like activity beyond the orbit of Jupiter. Our targets were selected from objects newly discovered in the \PSone\ detection database and thus have orbital arcs covering up to a decade. Most objects in our sample have orbits crossing the range of heliocentric distances where the temperatures allow for  crystallization of AWI speculated to be the major activity driver in Centaurs. We conclude that:
\begin{enumerate}
\item None of the Centaurs in our sample showed signs of activity either in our observations with Gemini, or in archival images from DECam dating back to 2013. However, we cannot rule out the presence of faint coma below our detction limits.
    
\item  The thermo-dynamical model we used allowed us to calculate the theoretical maximal sublimation rate of the three most common volatile species found on comets - CO, CO$_2$ and H$_2$O, and our results suggests the majority of sublimation-driven activity at the analyzed heliocentric distances should be due to CO. Due to the heliocentric distances and orbital histories of investigated Centaurs, it is unlikely that free CO or CO$_2$ remains on the surfaces, but it is possible our targets retain subsurface volatile deposits and some of them could exhibit low-level gas and dust mass-loss below our detection limits. 

\item  The upper limit on dust production rates and fractional active surface areas resulting from our coma modelling based on the non-detection of activity and the maximum surface sublimation rates are in agreement with rates estimated independently for other inactive Centaurs \citep[\eg][]{Bauer2003,Jewitt2009,Li2020}. If there are no surface super volatiles present and sublimating, the estimated dust gas production rates could be sustained by a modestly-sized (volume in order of several m$^2$) exposed pockets of AWI crystallizing and releasing trapped gas molecules.

\item  Our orbital history analysis revealed that most of our targets have relatively stable orbits with no significant recent changes in perihelion distance. Our results are consistent with earlier work suggesting that the activity on Centaurs is most likely triggered by sudden drops in perihelion distance and/or semi-major axis induced by close encounters with giant planets, which change the thermal balance in the body. The rapid warming and crystallization of AWI can then release trapped gas and build up pressure under the surface which leads to exposure of more volatile material and subsequent outbursts.

\item Our numerical integrations show that Centaur 2014 PQ$_{70}$ has experienced a drop in semi-major axis larger than 0.5 AU about 50 years ago, similar to what has been seen in other active Centaurs in the past, and is therefore a good candidate for monitoring of possible future outbursts of activity.

\end{enumerate}

\newpage
\acknowledgements
\section*{Acknowledgments}

This work was supported by NSF AST Grant \#1910275.

 We thank Kat Volk for helpful insights into Centaur dynamics and Yan Fern{\'a}ndez for insights into the dust-to-gas ratio of 29P.

The authors recognize and acknowledge the very significant cultural role and reverence that the summit of Mauna Kea has always had within the indigenous Hawaiian community. We are most fortunate to have the opportunity to conduct observations from this mountain.

This project used public archival data from the Dark Energy Survey (DES). Funding for the DES Projects has been provided by the U.S. Department of Energy, the U.S. National Science Foundation, the Ministry of Science and Education of Spain, the Science and Technology FacilitiesCouncil of the United Kingdom, the Higher Education Funding Council for England, the National Center for Supercomputing Applications at the University of Illinois at Urbana-Champaign, the Kavli Institute of Cosmological Physics at the University of Chicago, the Center for Cosmology and Astro-Particle Physics at the Ohio State University, the Mitchell Institute for Fundamental Physics and Astronomy at Texas A\&M University, Financiadora de Estudos e Projetos, Funda{\c c}{\~a}o Carlos Chagas Filho de Amparo {\`a} Pesquisa do Estado do Rio de Janeiro, Conselho Nacional de Desenvolvimento Cient{\'i}fico e Tecnol{\'o}gico and the Minist{\'e}rio da Ci{\^e}ncia, Tecnologia e Inova{\c c}{\~a}o, the Deutsche Forschungsgemeinschaft, and the Collaborating Institutions in the Dark Energy Survey. The Collaborating Institutions are Argonne National Laboratory, the University of California at Santa Cruz, the University of Cambridge, Centro de Investigaciones Energ{\'e}ticas, Medioambientales y Tecnol{\'o}gicas-Madrid, the University of Chicago, University College London, the DES-Brazil Consortium, the University of Edinburgh, the Eidgen{\"o}ssische Technische Hochschule (ETH) Z{\"u}rich,  Fermi National Accelerator Laboratory, the University of Illinois at Urbana-Champaign, the Institut de Ci{\`e}ncies de l'Espai (IEEC/CSIC), the Institut de F{\'i}sica d'Altes Energies, Lawrence Berkeley National Laboratory, the Ludwig-Maximilians Universit{\"a}t M{\"u}nchen and the associated Excellence Cluster Universe, the University of Michigan, the National Optical Astronomy Observatory, the University of Nottingham, The Ohio State University, the OzDES Membership Consortium, the University of Pennsylvania, the University of Portsmouth, SLAC National Accelerator Laboratory, Stanford University, the University of Sussex, and Texas A\&M University.

Based in part on observations at Cerro Tololo Inter-American Observatory, National Optical Astronomy Observatory, which is operated by the Association of Universities for Research in Astronomy (AURA) under a cooperative agreement with the National Science Foundation.

This research has made use of data and/or services provided by the International Astronomical Union's Minor Planet Center. 

This work has made use of data from the European Space Agency (ESA) mission Gaia (https://www.cosmos.esa.int/gaia), processed by the Gaia Data Processing and Analysis Consortium (DPAC, https://www.cosmos.esa.int/web/gaia/dpac/consortium). Funding for the DPAC has been provided by national institutions, in particular the institutions participating in the Gaia Multilateral Agreement.

Eva Lilly would like to thank her children Betty and Sammy for being great nappers so this paper could be written, and she wishes to express her gratitude to PSI for having an excellent support system for families with children.

\bibliography{Centaurs}
\bibliographystyle{icarus} 
\newpage


\begin{deluxetable}{lccccccccc}
\tablewidth{440pt} 
\tabletypesize{\footnotesize}
\tablecolumns{10}
\tablecaption{Observing geometry, time of observation and photometry.}
\tablehead{
\colhead{Name} & 
\colhead{UT Date} & 
\colhead{$\Delta^a$} & 
\colhead{r$^b$} & 
\colhead{$\alpha^c$} &
\colhead{FWHM} & 
\colhead{Airmass} & 
\colhead{$\Phi^d$} & 
\colhead{$m_r^e$} & 

}
\startdata
2013 MZ$_{11}$	&	2017 Sep 24	&	19.8	&	20.6	&	1.8	&	0.76	&	1.75	&	4.74	&	21.24	$\pm$	0.02	\\
2014 NX$_{65}$	&	2017 Sep 24	&	17.7	&	18.7	&	0.8	&	0.62	&	1.21	&	3.37	&	22.00	$\pm$	0.23	\\
2014 TV$_{85}$	&	2017 Dec 11	&	14.6	&	14.9	&	3.5	&	0.67	&	1.07	&	4.05	&	21.26	$\pm$	0.02	\\
2015 BD$_{518}$	&	2017 Nov 14	&	16.5	&	16.3	&	3.4	&	0.73	&	1.31	&	2.00	&	21.93	$\pm$	0.03	\\
2015 BH$_{518}$	&	2017 Dec 30	&	25.1	&	25.4	&	2.1	&	0.76	&	1.15	&	4.05	&	22.60	$\pm$	0.07	\\
2014 WW$_{508}$	&	2017 Nov 17	&	13.3	&	14.0	&	3.4	&	0.67	&	1.02	&	3.37	&	22.17	$\pm$	0.03	\\
2010 TU$_{191}$	&	2018 Jan 16	&	14.8	&	15.5	&	2.1	&	1.05	&	1.31	&	3.37	&	22.81	$\pm$	0.04	\\
2014 JD$_{80}$	&	2017 Sep 19	&	19.2	&	19.8	&	2.2	&	1.54	&	1.70	&	4.00	&	21.83	$\pm$	0.04	\\
2014 OX$_{393}$	&	2017 Nov 9	&	11.3	&	12.3	&	0.6	&	0.69	&	1.48	&	4.05	&	21.63	$\pm$	0.02	\\
2014 YX$_{49}$	&	2017 Nov 11	&	19.3	&	19.6	&	2.7	&	0.62	&	1.04	&	2.50	&	21.66	$\pm$	0.02	\\
2014 WX$_{508}$	&	2017 Dec 11	&	13.4	&	13.4	&	3.9	&	0.63	&	1.16	&	4.05	&	22.32	$\pm$	0.10	\\
2015 BK$_{518}$	&	2018 Jan 17	&	7.9	    &	8.1	    &	6.8	&	1.13	&	1.25	&	8.16	&	23.71	$\pm$	0.12	\\
2014 PQ$_{70}$	&	2017 Sep 23	&	7.2	    &	8.1	    & 	6.6	&	1.11	&	1.39	&	7.47	&	24.02	$\pm$	0.14	\\
\enddata
\tablenotetext{*}{\textbf{Notes.}\\
$^a$ Geocentric distance, AU.\\
$^b$ Heliocentric distance, AU.\\
$^c$ Solar phase angle, deg.\\
$^d$ Angular diameter of photometry aperture, arcseconds.\\
$^e$ Measured mean apparent r'-band magnitude.\\
}
\label{tab.ObsConditions}
\end{deluxetable}

\begin{deluxetable}{lccccccccc}
\tablewidth{360pt} 
\tabletypesize{\footnotesize}
\tablecolumns{9}
\tablecaption{Orbital parameters.}
\tablehead{
\colhead{Name} & 
\colhead{q$^a$} & 
\colhead{a$^b$} & 
\colhead{e$^c$} & 
\colhead{i$^d$} &
\colhead{M$^e$} & 
\colhead{$T_J^f$} & 
\colhead{$T_S^g$} & 
\colhead{$T_U^h$} & 
}
\startdata
2013 MZ$_{11}$	&	16.828	&	24.228	&	0.305	&	6.40	         &	314.40	&	4.30	&	3.41	&	2.92	\\
2014 NX$_{65}$	&	18.386	&	22.868	&	0.196	&	11.40	&	16.30	&	4.26	&	3.39	&	2.94	\\
2014 TV$_{85}$	&	14.653	&	21.796	&	0.328	&	12.20	&	348.40	&	4.02	&	3.22	&	2.85	\\
2015 BD$_{518}$	&	16.277	&	23.383	&	0.304	&	17.20	&	359.20	&	4.08	&	3.25	&	2.83	\\
2015 BH$_{518}$	&	25.356	&	27.954	&	0.093	&	10.90	&	9.70	        &	4.72	&	3.68	&	3.05	\\
2014 WW$_{508}$	&	13.259	&	28.219	&	0.53	        &	9.50	         &	8.20	        &	4.08	&	3.21	&	2.71	\\
2010 TU$_{191}$	&	14.764	&	20.65	&	0.285	&	1.80	         &	19.10	&	4.07	&	3.28	&	2.92	\\
2014 JD$_{80}$	&	19.012	&	25.318	&	0.249	&	39.10	&	337.90	&	3.52	&	2.82	&	2.48	\\
2014 OX$_{393}$	&	11.963	&	25.444	&	0.53	        &	14.70	&	5.60	        &	3.83	&	3.05	&	2.64	\\
2014 YX$_{49}$	&	13.825	&	19.113	&	0.277	&	25.60	&	78.10	&	3.59	&	2.95	&	2.73	\\
2014 WX$_{508}$	&	13.173	&	18.631	&	0.293	&	11.60	&	348.90	&	3.82	&	3.13	&	2.88	\\
2015 BK$_{518}$	&	6.393	&	14.258	&	0.552	&	7.00	        &	16.50	&	3.10	&	2.69	&	2.77	\\
2014 PQ$_{70}$	&	6.25443	&	16.037	&	0.6	         &	15.04	&	11.80	&	3.01	&	2.58	&	2.60	\\
\enddata

\tablenotetext{*}{\textbf{Notes.}\\
$^a$ Perihelion distance, AU.\\
$^b$ Semimajor axis, AU.\\
$^c$ Orbital eccentricity\\
$^d$ Inclination, degrees.\\
$^e$ Mean anomaly at the time of observation, degrees.\\
$^f$ Tisserand'sparameter with respect to Jupiter.\\
$^g$ Tisserand's parameter with respect to Saturn.\\
$^h$ Tisserand's parameter with respect to Uranus.\\
}
\label{tab.Orbparameters}
\end{deluxetable}

\begin{deluxetable}{lccclllcllc}
\tablewidth{500pt} 
\tabletypesize{\scriptsize}
\tablecolumns{11}
\tablecaption{Upper Limits on Sublimation and Dust Production Rates}
\tablehead{
\colhead{Name} & 
\colhead{$H_r^{\textcolor{blue}{a}}$} & 
\colhead{$R^{\textcolor{blue}{b}}$} & 
\colhead{$\eta^{\textcolor{blue}{c}}$} & 
\colhead{$dM^{\textcolor{blue}{d}}_{H_2O}/dt$} &
\colhead{$dM^{\textcolor{blue}{e}}_{CO_2}/dt$} &
\colhead{$dM^{\textcolor{blue}{f}}_{CO}/dt$} &
\colhead{$dM^{\textcolor{blue}{g}}_D/dt$} & 
\colhead{$f_{Act}^{\textcolor{blue}{h}}$} & 
\colhead{$A_{Act}^{\textcolor{blue}{i}}$} & 
\colhead{$V_{AWI}^{\textcolor{blue}{j}}$}
}
\startdata
2013 MZ$_{11}$	&	7.95	&	60.29	&	0.25	&	$2.54 \times 10^{-25}$	&	$1.74 \times 10^{-2}$	&	$1.27 \times 10^7$	&	10.80	&	$8.50 \times {-7}$	&	$3.88 \times 10^4$	&	27.14	\\
2014 NX$_{65}$	&	9.27	&	32.91	&	0.05	&	$9.84 \times 10^{-24}$	&	$4.82 \times 10^{-2}$	&	$3.57 \times 10^6$	&	0.76	&	$2.10 \times {-7}$	&	$2.89 \times 10^3$	&	11.19	\\
2014 TV$_{85}$	&	9.23	&	33.49	&	0.10	&	$6.04 \times 10^{-19}$	&	$9.77 \times 10^0$	&	$5.23 \times 10^6$	&	2.14	&	$4.09 \times {-7}$	&	$5.76 \times 10^3$	&	15.81	\\
2015 BD$_{518}$	&	9.45	&	30.30	&	0.05	&	$8.66 \times 10^{-21}$	&	$1.24\times 10^0$	&	$4.34 \times 10^6$	&	0.74	&	$1.70 \times {-7}$	&	$1.96 \times 10^3$	&	11.10	\\
2015 BH$_{518}$	&	8.32	&	50.88	&	0.05	&	$1.44 \times 10^{-30}$	&	$4.69 \times 10^{-5}$	&	$7.39 \times 10^6$	&	1.09	&	$1.48 \times {-7}$	&	$4.81 \times 10^3$	&	12.65	\\
2014 WW$_{508}$	&	10.48	&	18.82	&	0.05	&	$4.49 \times 10^{-18}$	&	$1.69 \times 10^1$	&	$2.60 \times 10^6$	&	0.38	&	$1.47 \times {-7}$	&	$6.53 \times10^2$	&	8.90	\\
2010 TU$_{191}$	&	10.75	&	16.61	&	0.05	&	$3.81 \times 10^{-20}$	&	$1.59 \times 10^0$	&	$1.92 \times 10^6$	&	0.25	&	$1.32 \times {-7}$	&	$4.57 \times 10^2$	&	7.77	\\
2014 JD$_{80}$	&	8.68	&	43.20	&	0.05	&	$7.76 \times 10^{-25}$	&	$1.78 \times 10^{-2}$	&	$5.02 \times 10^6$	&	1.17	&	$2.33 \times {-7}$	&	$5.48 \times 10^3$	&	12.94	\\
2014 OX$_{393}$	&	10.79	&	16.31	&	0.05	&	$8.56 \times 10^{-16}$	&	$1.78 \times 10^2$	&	$2.34 \times 10^6$	&	0.36	&	$1.53 \times 10^{-7}$	&	$5.13 \times 10^2$	&	8.72	\\
2014 YX$_{49}$	&	8.49	&	47.10	&	0.05	&	$1.92 \times 10^{-24}$	&	$3.34 \times 10^{-2}$	&	$7.41 \times 10^6$	&	1.40	&	$1.88 \times 10^{-7}$	&	$5.26 \times 10^3$	&	13.72	\\
2014 WX$_{508}$	&	10.68	&	17.15	&	0.05	&	$2.60 \times 10^{-17}$	&	$3.61 \times 10^1$	&	$2.36 \times 10^6$	&	0.32	&	$1.36 \times 10^{-7}$	&	$5.02 \times 10^2$	&	8.40	\\
2015 BK$_{518}$	&	14.16	&	3.46	&	0.10	&	$3.15 \times 10^{-10}$	&	$4.65 \times 10^3$	&	$2.64 \times 10^5$	&	0.06	&	$2.17 \times 10^{-7}$	&	$3.27 \times 10^1$	&	4.73	\\
2014 PQ$_{70}$	&	14.71	&	2.69	&	0.10	&	$3.73 \times 10^{-10}$	&	$5.51 \times 10^3$	&	$3.13 \times 10^5$	&	0.04	&	$1.21 \times 10^{-7}$	&	$1.10 \times 10^1$	&	4.13	\\
\enddata
\tablenotetext{*}{\textbf{Notes.}\\
$^{\textcolor{blue}{a}}$ Absolute R-band total magnitude (at R=$\Delta$=1 AU and $\alpha$= 0 deg), using IAU H, G phase-darkening, where G = 0.13 \citep{Bauer2003}.\\
$^{\textcolor{blue}{b}}$ Effective nucleus radius, km.\\
$^{\textcolor{blue}{c}}$ Coma limit parameter.\\
$^{\textcolor{blue}{d}}$ Maximum H$_2$O-sublimation rate, in kg~s$^{-1}$.\\
$^{\textcolor{blue}{e}}$ Maximum CO$_2$-sublimation rate, in kg~s$^{-1}$.\\
$^{\textcolor{blue}{f}}$ Maximum CO-sublimation rate, in kg~s$^{-1}$.\\
$^{\textcolor{blue}{g}}$ CO-driven dust production rate, calculated assuming $f_{dg}=6$, in kg~s$^{-1}$.\\
$^{\textcolor{blue}{h}}$ Effective active surface fraction of CO-driven activity. \\
$^{\textcolor{blue}{i}}$ Effective active surface area of CO-driven activity, in $m^2$. \\
$^{\textcolor{blue}{j}}$ Effective radius of a spherical AWI volume if solely sustaining upper limit on the CO gas flow, in $m$. \\
}
\label{tab.Mag_radii_production}
\end{deluxetable}

\begin{deluxetable}{lccccccccc}
\tablewidth{460pt} 
\tabletypesize{\footnotesize}
\tablecolumns{9}
\tablecaption{DECam archival detections of or Centaur targets.}
\tablehead{
\colhead{Name} & 
\colhead{UT Date} & 
\colhead{UT Time} & 
\colhead{Filter} &
\colhead{$\Delta^a$} & 
\colhead{r$^b$} & 
\colhead{$\alpha^c$} & 
\colhead{Airmass} & 
\colhead{$m^d$} & 
}
\startdata
2013 MZ$_{11}$ & 2019-08-10 & 05:24:31.443 & i$'$ & 18.891 & 19.874 & 0.732 & 1.04 & 20.87 $\pm$ 0.04 \\
2013 MZ$_{11}$ & 2019-09-25 & 00:48:04.371 & VR & 18.976 & 19.826 & 1.572 & 1.15 & 21.38 $\pm$ 0.01 \\
2014 JD$_{80}$ & 2015-08-16 & 02:54:48.666 & VR & 19.315 & 20.24 & 1.193 & 1.01 & 22.02 $\pm$ 0.13 \\
2014 JD$_{80}$ & 2015-08-16 & 04:33:52.183 & VR & 19.316 & 20.24 & 1.196 & 1.03 & 22.05 $\pm$ 0.29 \\
2014 JD$_{80}$ & 2015-09-05 & 23:47:24.726 & VR & 19.493 & 20.227 & 1.993 & 1.14 & 22.39 $\pm$ 0.10 \\
2014 JD$_{80}$ & 2017-06-03 & 06:09:40.435 & i$'$ & 19.311 & 19.854 & 2.509 & 1.26 & 22.52 $\pm$ 0.34 \\
2014 JD$_{80}$ & 2017-07-02 & 07:23:43.929 & g$'$ & 18.975 & 19.838 & 1.588 & 1.0 & 22.16 $\pm$ 0.32 \\
2014 JD$_{80}$ & 2017-07-17 & 07:32:02.319 & i$'$ & 18.878 & 19.83 & 1.058 & 1.05 & 21.56 $\pm$ 0.10 \\
2014 JD$_{80}$ & 2017-08-27 & 02:08:09.002 & r$'$ & 18.932 & 19.808 & 1.488 & 1.04 & 21.33 $\pm$ 0.14 \\
2014 JD$_{80}$ & 2017-08-27 & 05:28:54.472 & r$'$ & 18.933 & 19.808 & 1.493 & 1.13 & 21.01 $\pm$ 0.10 \\
2014 JD$_{80}$ & 2017-08-28 & 01:49:49.905 & r$'$ & 18.939 & 19.808 & 1.524 & 1.05 & 21.04 $\pm$ 0.13 \\
2014 JD$_{80}$ & 2017-09-01 & 02:50:26.376 & r$'$ & 18.97 & 19.806 & 1.673 & 1.0 & 21.65 $\pm$ 0.24 \\
2014 JD$_{80}$ & 2017-09-01 & 03:51:49.346 & r$'$ & 18.971 & 19.806 & 1.674 & 1.02 & 21.36 $\pm$ 0.20 \\
2014 NX$_{65}$ & 2017-07-30 & 06:35:06.476 & g$'$ & 17.94 & 18.65 & 2.275 & 1.12 & 23.34 $\pm$ 0.81 \\
2014 NX$_{65}$ & 2017-07-30 & 06:36:37.613 & g$'$ & 17.94 & 18.65 & 2.275 & 1.11 & 23.05 $\pm$ 0.32 \\
2014 NX$_{65}$ & 2017-08-20 & 04:34:21.015 & r$'$ & 17.751 & 18.656 & 1.43 & 1.19 & 22.41 $\pm$ 0.22 \\
2014 NX$_{65}$ & 2018-08-12 & 05:09:17.383 & r$'$ & 17.981 & 18.771 & 1.982 & 1.26 & 22.36 $\pm$ 0.50 \\
2014 NX$_{65}$ & 2018-09-07 & 03:36:18.323 & g$'$ & 17.814 & 18.78 & 0.902 & 1.22 & 23.58 $\pm$ 0.24 \\
2014 NX$_{65}$ & 2018-09-12 & 03:13:44.044 & r$'$ & 17.803 & 18.782 & 0.737 & 1.23 & 22.02 $\pm$ 0.16 \\
2014 NX$_{65}$ & 2018-09-12 & 03:15:04.320 & g$'$ & 17.803 & 18.782 & 0.737 & 1.22 & 23.17 $\pm$ 0.28 \\
2014 NX$_{65}$ & 2019-09-26 & 05:30:12.957 & VR & 17.945 & 18.927 & 0.634 & 1.08 & 22.21 $\pm$ 0.03 \\
2014 OX$_{393}$ & 2014-10-03 & 07:06:38.523 & g$'$ & 11.047 & 12.032 & 0.863 & 1.34 & 23.04 $\pm$ 0.41 \\
2014 OX$_{393}$ & 2014-10-03 & 07:08:36.795 & r$'$ & 11.047 & 12.032 & 0.863 & 1.35 & 21.80 $\pm$ 0.11 \\
2014 OX$_{393}$ & 2014-10-03 & 07:10:35.396 & i$'$ & 11.047 & 12.032 & 0.863 & 1.36 & 21.87 $\pm$ 0.10 \\
2014 OX$_{393}$ & 2015-11-02 & 02:53:11.989 & r$'$ & 11.05 & 11.969 & 1.87 & 1.16 & 20.38 $\pm$ 0.16 \\
2014 OX$_{393}$ & 2015-12-28 & 01:43:08.750 & i$'$ & 11.764 & 11.974 & 4.637 & 1.4 & 21.50 $\pm$ 0.15 \\
2014 OX$_{393}$ & 2015-12-28 & 03:00:59.051 & i$'$ & 11.765 & 11.974 & 4.638 & 1.94 & 21.84 $\pm$ 0.15 \\
2014 OX$_{393}$ & 2016-07-22 & 09:53:40.451 & z$'$ & 12.005 & 12.022 & 4.848 & 1.29 & 21.77 $\pm$ 0.10 \\
2014 OX$_{393}$ & 2016-07-22 & 10:10:05.819 & z$'$ & 12.005 & 12.022 & 4.848 & 1.27 & 21.25 $\pm$ 0.10 \\
2014 OX$_{393}$ & 2017-11-13 & 04:18:13.011 & r$'$ & 11.337 & 12.306 & 0.963 & 1.39 & 21.80 $\pm$ 0.14 \\
2014 TV$_{85}$ & 2015-03-09 & 04:16:42.821 & VR & 14.884 & 15.59 & 2.626 & 1.68 & 21.83 $\pm$ 0.06 \\
2014 TV$_{85}$ & 2015-04-11 & 23:29:44.974 & z$'$ & 15.345 & 15.562 & 3.627 & 1.31 & 21.41 $\pm$ 0.11 \\
2014 TV$_{85}$ & 2016-01-10 & 05:46:06.736 & r$'$ & 14.455 & 15.349 & 1.582 & 1.26 & 21.29 $\pm$ 0.08 \\
2014 TV$_{85}$ & 2016-01-10 & 05:47:43.758 & g$'$ & 14.455 & 15.349 & 1.582 & 1.26 & 21.75 $\pm$ 0.13 \\
2014 TV$_{85}$ & 2017-03-01 & 00:50:23.512 & r$'$ & 14.173 & 15.073 & 1.622 & 1.51 & 21.17 $\pm$ 0.08 \\
2014 TV$_{85}$ & 2017-03-01 & 00:55:23.608 & g$'$ & 14.173 & 15.073 & 1.622 & 1.49 & 21.63 $\pm$ 0.10 \\
2014 TV$_{85}$ & 2018-01-28 & 06:50:34.844 & z$'$ & 13.981 & 14.898 & 1.431 & 1.21 & 20.88 $\pm$ 0.06 \\
2014 TV$_{85}$ & 2018-02-05 & 06:19:29.275 & i$'$ & 13.943 & 14.894 & 1.037 & 1.21 & 20.69 $\pm$ 0.08 \\
2014 TV$_{85}$ & 2018-12-13 & 06:21:48.624 & g$'$ & 14.489 & 14.772 & 3.694 & 1.64 & 21.62 $\pm$ 0.11 \\
2014 WW$_{508}$ & 2016-03-03 & 01:15:05.013 & VR & 13.159 & 13.47 & 4.051 & 1.49 & 21.10 $\pm$ 0.13 \\
2014 WW$_{508}$ & 2019-02-28 & 02:32:20.617 & VR & 13.839 & 14.57 & 2.697 & 1.38 & 22.34 $\pm$ 0.15 \\
2014 WW$_{508}$ & 2019-02-28 & 02:35:19.173 & VR & 13.839 & 14.57 & 2.697 & 1.39 & 22.55 $\pm$ 0.22 \\
2014 YX$_{49}$ & 2016-12-29 & 06:27:42.080 & g$'$ & 18.301 & 19.21 & 1.151 & 2.16 & 21.97 $\pm$ 0.11 \\
2015 BD$_{518}$ & 2014-12-29 & 08:26:36.064 & z$'$ & 15.716 & 16.466 & 2.264 & 1.73 & 21.62 $\pm$ 0.13 \\
2015 BD$_{518}$ & 2015-04-12 & 00:52:43.348 & r$'$ & 16.026 & 16.432 & 3.233 & 1.72 & 22.30 $\pm$ 0.14 \\
2015 BD$_{518}$ & 2015-04-12 & 00:54:11.319 & g$'$ & 16.026 & 16.432 & 3.234 & 1.72 & 22.48 $\pm$ 0.15 \\
2015 BD$_{518}$ & 2016-01-15 & 06:58:55.054 & r$'$ & 15.498 & 16.356 & 1.729 & 1.56 & 22.98 $\pm$ 0.25 \\
2015 BD$_{518}$ & 2016-01-15 & 07:00:25.880 & g$'$ & 15.498 & 16.356 & 1.729 & 1.56 & 23.12 $\pm$ 0.23 \\
2015 BD$_{518}$ & 2017-01-11 & 05:59:11.086 & z$'$ & 15.523 & 16.295 & 2.198 & 1.62 & 21.51 $\pm$ 0.06 \\
2015 BD$_{518}$ & 2017-03-30 & 01:02:15.573 & g$'$ & 15.539 & 16.287 & 2.381 & 1.57 & 22.01 $\pm$ 0.13 \\
2015 BD$_{518}$ & 2018-02-12 & 05:36:58.517 & g$'$ & 15.312 & 16.277 & 0.746 & 1.36 & 22.16 $\pm$ 0.07 \\
2015 BH$_{518}$ & 2013-04-01 & 02:24:49.422 & z$'$ & 24.537 & 25.345 & 1.351 & 1.14 & 19.46 $\pm$ 0.10 \\
2015 BH$_{518}$ & 2014-03-28 & 00:03:22.996 & VR & 24.479 & 25.35 & 1.118 & 1.49 & 22.12 $\pm$ 0.11 \\
2015 BH$_{518}$ & 2015-03-29 & 23:42:12.581 & z$'$ & 24.478 & 25.36 & 1.077 & 1.63 & 22.38 $\pm$ 0.11 \\
2015 BK$_{518}$ & 2017-03-02 & 05:30:37.098 & g$'$ & 6.34 & 7.307 & 1.826 & 1.11 & 23.30 $\pm$ 0.24 \\
\enddata
\tablenotetext{*}{\textbf{Notes.}\\
$^a$ Geocentric distance at the time of observation, AU.\\
$^b$ Heliocentric distance at the time of observation, AU.\\
$^c$ Phase angle at the time of observation.\\
$^d$ Apparent magnitude measured in given filter.\\
}

\label{tab.DECam_obs}
\end{deluxetable}

\newpage

\end{document}